\documentclass[preprint,10pt]{elsarticle}

\journal{Elsevier}

\usepackage{graphicx,bbm,setspace,amssymb,amsfonts,amsmath,mathtools,mathrsfs,stmaryrd,amsthm,bm,etoolbox,comment,hyperref,url,caption,subcaption,color,enumitem,algorithm,booktabs,microtype,nicefrac,lingmacros,nccmath,tree-dvips,tikz-cd, longtable}
\usepackage[noend]{algpseudocode}
\usepackage{algorithm}

\usepackage[colorinlistoftodos]{todonotes}
\usepackage[utf8]{inputenc} 
\usepackage[T1]{fontenc}    

\usepackage[noend]{algpseudocode}

\newtheorem{thm-defn}[theorem]{Theorem/Definition}

\theoremstyle{definition}

\theoremstyle{remark}

\usepackage{amssymb,amsmath,mathrsfs,stmaryrd,amsthm,mathtools,amsfonts,graphicx,hyperref,url,booktabs,nicefrac,comment,microtype,tikz-cd,lingmacros,nccmath,tree-dvips, bbm, bm, etoolbox,algorithm,caption,subcaption,color,enumitem, physics, lineno, setspace} 

\DeclareMathOperator*{\argmax}{argmax}

\newcommand{\bs}[1]{\bm{#1}} 

\newcommand{\ignore}[1]{}{}

\begin{document}

\begin{frontmatter}

\title{Evolutionary correlation, regime switching, spectral dynamics and optimal trading strategies for cryptocurrencies and equities}
   
\author[label1]{Nick James} \ead{nick.james@unimelb.edu.au}
\address[label1]{School of Mathematics and Statistics, University of Melbourne, Victoria, Australia}

\begin{abstract}

This paper uses new and recently established methodologies to study the evolutionary dynamics of the cryptocurrency market, and compares the findings with that of the equity market. We begin by applying random matrix theory and principal components analysis (PCA) to correlation matrices of both collections, highlighting clear differences in the eigenspectra exhibited. We then explore the heterogeneity of both asset classes, studying the time-varying dynamics of underlying sector behaviours, and determine the collective similarity within each collection. We then turn to a study of structural break dynamics and evolutionary power spectra, where we quantify the collective affinity in structural breaks and evolutionary behaviours of underlying sector time series. Finally, we implement two algorithms simulating `portfolio choice' dynamics to compare the effectiveness of stock selection and sector allocation in cryptocurrency portfolios. There, we highlight the importance of both endeavours and comment on noteworthy implications for cryptocurrency portfolio management.

\end{abstract}


\end{frontmatter}

\section{Introduction}
\label{Introduction}

Since its inception, the cryptocurrency market has seen rapid growth in its total market capitalisation and gradual changes in its underlying dynamics. The increase in the cryptocurrency market's maturity has been accompanied by rapid growth in the number of cryptocurrencies available to trade, and a general rise in the level of sophistication across the asset class. In this paper, we seek to explore the evolution of the cryptocurrency market, and compare its behaviours to the larger, and more sophisticated equity market.

A variety of communities have been interested in the study of financial market dynamics, and the evolutionary changes in structural behaviours, including researchers in applied mathematics, econometrics and econophysics \citep{Fenn2011, Laloux1999, Mnnix2012}. A variety of mathematical techniques have been used to analyze the evolutionary dynamics exhibited by such financial assets including principal components analysis (PCA) \cite{Laloux1999, Kim2005, Pan2007, Wilcox2007}, random matrix theory \cite{Miceli2004, Conlon2007, Bouchaud2003, Burda2004, Sharifi2004, Fenn2011}, various types of clustering \cite{Heckens2020, Jamesfincovid}, structural break detection \cite{James2021_crypto, JamescryptoEq} and a variety of mathematical and statistical modelling techniques \cite{Tilfani2022, Ferreira2020}. These techniques have been used on a wide variety of asset classes such as equities \cite{Wilcox2007}, foreign exchange \cite{Ausloos2000} and fixed income \cite{Driessen2003}. In more recent years, researchers have explored the application of such techniques as a means of understanding the burgeoning cryptocurrency market. Various aspects of the market have been studied including time-varying behaviours \cite{James2021_crypto, JamescryptoEq, Kwapie2021, Wtorek2020, Drod2018, Drod2019, Drod2020, Drod2020_entropy}, Bitcoin's price behaviours \cite{Chu2015, Lahmiri2018, Kondor2014, Bariviera2017, AlvarezRamirez2018}, fractal-type dynamics and pattern formation, \cite{Stosic2019, Stosic2019_2, Manavi2020, Ferreira2020}, and trading strategies \cite{Fil2020}.

The study of nonstationarity behaviours and the time-varying nature of (autoregressive) model parameters for phenomena such as volatility, has been of great interest to financial markets researchers for many years. Such research dates back to Autoregressive conditionally heteroskedastic (ARCH) models \cite{Engle1982}, Generalized ARCH (GARCH) \cite{Bollerslev1986} and stochastic volatility models \cite{Taylor1982, Taylor1986, Taylor1994}. In more recent years there have been various extensions and adaptations to the core models, allowing for the specific consideration of various time series features. A non-exhaustive collection of model extensions may mention EGARCH \cite{Nelson1991}, GJR-GARCH \cite{GLOSTEN1993}, T-GARCH \cite{Zakoian1994} and T-SV \cite{So2002}, Markov Switching GARCH \cite{Cai1994, Hamilton1994, Gray1996} and MS-SV \cite{Lam1998}. Bayesian estimation has also been prominent in this area \cite{Andersen2003, Hwang2007, Hansen2011}. Such modelling procedures and various adaptations have been widely applied to study the properties of cryptocurrencies and other asset classes \cite{Cerqueti2020, Wan2017, Stehlk2017, ChiaShangJamesChu1996, Chen2018, Cerqueti2019}. In this work specifically, we build on the concept of local stationarity in stochastic processes \cite{Dahlhaus1997, Adak1998} and time-varying spectral density estimation \cite{Rosen2009, Rosen2012, Rosen2017, james2021_spectral}, to study the propensity for regime switching, and the associated changes in periodic behaviour, in cryptocurrency and equity sectors.

Portfolio optimization, and more generally the concept of portfolio construction, has been a topic of considerable interest to the mathematics and quantitative finance communities \cite{Markowitz1952, Sharpe1966, Almahdi2017, Calvo2014, Soleimani2009, Vercher2007, Bhansali2007, Moody2001}. Although there is literature exploring the effectiveness of cryptocurrencies as safe-haven portfolio assets, we are unaware of any work explicitly testing for comparative effectiveness in alpha generation when contrasting security selection, and sector allocation. In this work, we introduce two algorithmic procedures to test this hypothesis in the cryptocurrency market.

This paper is structured as follows. In Section \ref{RMT} we apply random matrix theory in stationary and time-varying contexts, to reveal differences in the impact of the `market factor' and other factors on the correlations of both collections' security returns. In Section \ref{Evolutionary_dynamics_sector} we extend our study of asset class heterogeneity, and compare the evolution in collective strength among underlying cryptocurrency and equity sectors. In Section \ref{Structural_breaks_non_stationarity} we contrast the similarity in underlying sectors' structural break dynamics, and further distinguish between nonstationarity and long-range dependence by quantifying the distance between representative time series' time-varying power spectra. Finally in Section \ref{Cryptocurrency_time_varying_optimization} we introduce two algorithms to determine the relative importance of security selection and sector allocation when assembling cryptocurrency portfolios. In Section \ref{Conclusion} we conclude.


\section{Data}
\label{Data}

In this paper we study cryptocurrency and equity prices between 01-01-2019 to 01-12-2021. We study a collection of 45 unique cryptocurrency securities and 77 unique equity securities. Cryptocurrencies are partitioned into 8 sectors (Centralized exchange, Collectibles/NFTs, Decentralized finance, Platform, Smart contracts, Store of value, Wallet) and equities are partitioned into 11 sectors (Communication services, Consumer discretionary, Consumer staples, Energy, Financials, Healthcare, Industrials, Information technology, Materials, Real estate, Utilities). Equity securities are classified based on their Global Industry Classification Standard (GICS) sector. Cryptocurrency data is sourced from Coinmarketcap.com (https://coinmarketcap.com/) and equity data is sourced from Yahoo Finance (https://au.finance.yahoo.com/). Some cryptocurrencies may be classified in more than one sector (as is consistent with the Coinmarketcap.com classification process).

\section{Time-varying random matrix theory and spectral dynamics}
\label{RMT}

\begin{figure}
    \centering
    \begin{subfigure}[b]{0.48\textwidth}
        \includegraphics[width=\textwidth]{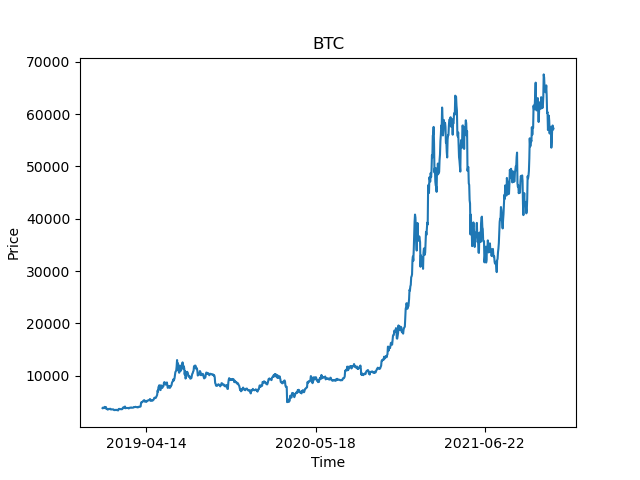}
        \caption{}
        \label{fig:Bitcoin_price}
    \end{subfigure}
    \begin{subfigure}[b]{0.48\textwidth}
        \includegraphics[width=\textwidth]{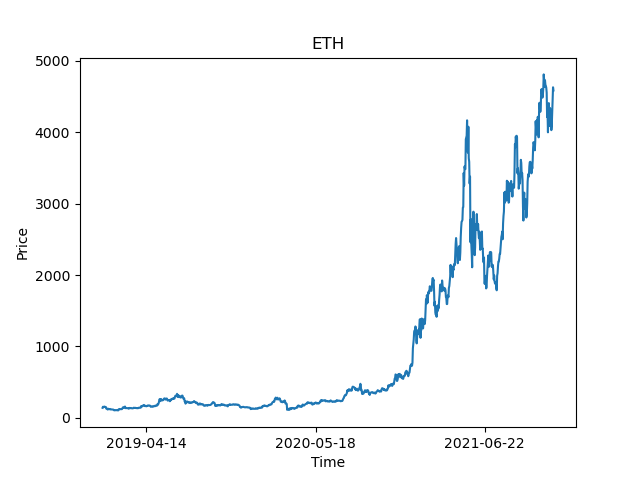}
        \caption{}
        \label{fig:Ethereum_price}
    \end{subfigure}
    \begin{subfigure}[b]{0.48\textwidth}
        \includegraphics[width=\textwidth]{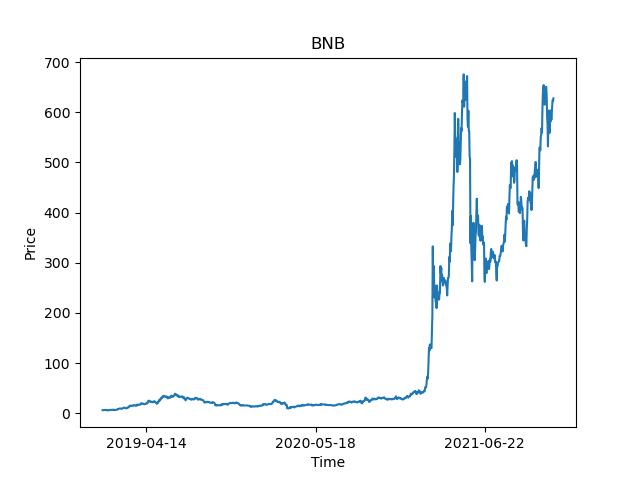}
        \caption{}
        \label{fig:BNB_price}
    \end{subfigure}
    \begin{subfigure}[b]{0.48\textwidth}
        \includegraphics[width=\textwidth]{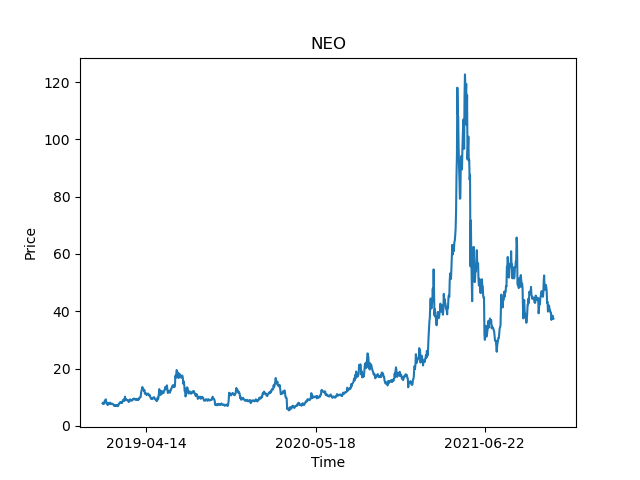}
        \caption{}
        \label{fig:Neo_price}
    \end{subfigure}
    \caption{A sample of cryptocurrency price time series for (a) Bitcoin (b) Ethereum, (c) Binance Coin, (d) NEO.}
    \label{fig:Cryptocurrency_time_series}
\end{figure}

\begin{figure}
    \centering
    \begin{subfigure}[b]{0.48\textwidth}
        \includegraphics[width=\textwidth]{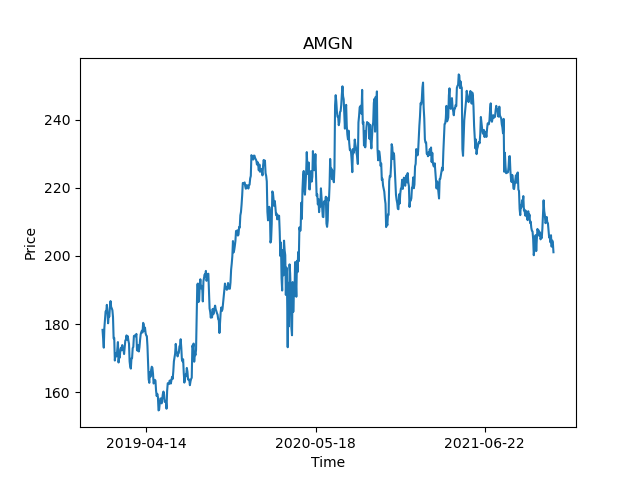}
        \caption{}
        \label{fig:Amgen_price}
    \end{subfigure}
    \begin{subfigure}[b]{0.48\textwidth}
        \includegraphics[width=\textwidth]{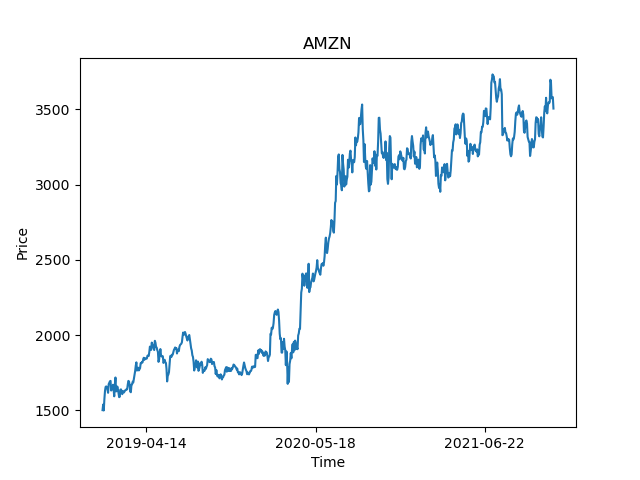}
        \caption{}
        \label{fig:Amazon_price}
    \end{subfigure}
    \begin{subfigure}[b]{0.48\textwidth}
        \includegraphics[width=\textwidth]{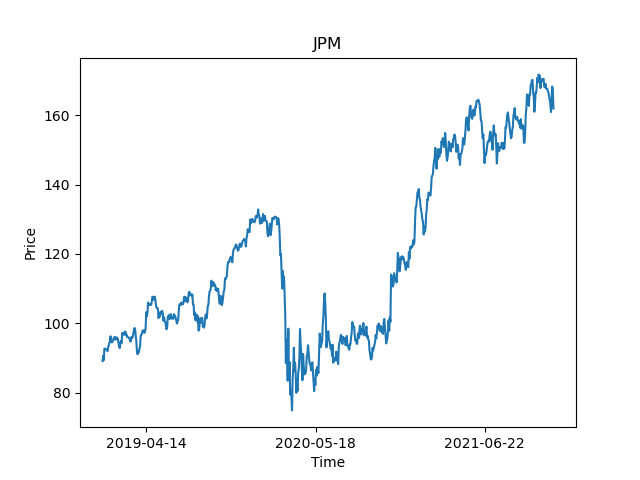}
        \caption{}
        \label{fig:JPM_price}
    \end{subfigure}
    \begin{subfigure}[b]{0.48\textwidth}
        \includegraphics[width=\textwidth]{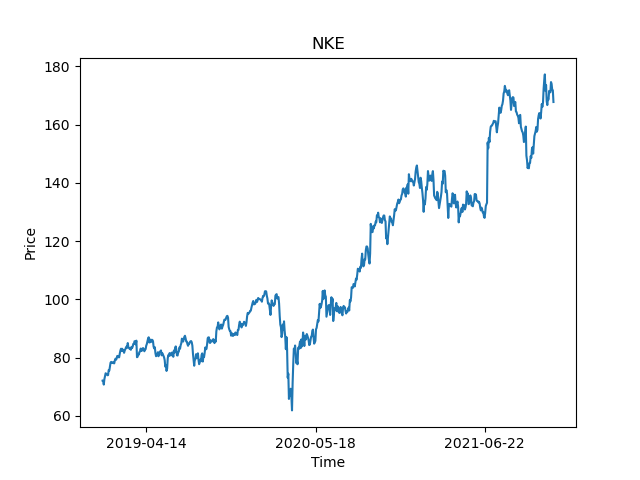}
        \caption{}
        \label{fig:NKE_price}
    \end{subfigure}
    \caption{A sample of equity price time series for (a) Amgen (b) Amazon, (c) JP Morgan, (d) Nike.}
    \label{fig:Equity_time_series}
\end{figure}

Throughout this paper we deal with two central objects of study, a collection of cryptocurrency time series and a collection of equity time series, where the underlying securities are uniformly sampled from a variety of respective sectors (see \ref{appendix:Securities} for the complete list of securities). Given that cryptocurrencies trade 7 days per week (unlike equities), although both collections span 01-01-2019 to 01-12-2021, the length of time series within each collection will be different. Let our cryptocurrency collection be indexed $t_1= 1,...,T_1$ where $T_1 = 1065$ and our equity collection be indexed $t_2= 1,...,T_2$ where $T_2 = 733$. Let $p^c_i(t_1)$ and $p^e_j(t_2)$ refer to the multivariate time series of cryptocurrency daily closing prices and equity adjusted daily closing prices respectively $i = 1,...,N$ and $j = 1,...,K$. A collection of representative price time series for our cryptocurrency and equity collections are shown in Figures \ref{fig:Cryptocurrency_time_series} and \ref{fig:Equity_time_series} respectively. We generate two multivariate time series of log returns $R^c_i(t_1)$ and $R^e_j(t_2)$ as follows:
\begin{align}
R^c_{i}{(t_1)} &= \log \left(\frac{p^c_i{(t_1)}}{p^c_i{(t_1-1)}}\right), \\
R^e_{j}{(t_2)} &= \log \left(\frac{p^e_j{(t_2)}}{p^e_j{(t_2-1)}}\right).
\end{align}

We standardise cryptocurrency and equity log returns as follows. For our cryptocurrency collection, let $\tilde{R}^c_i(t_1) = [R^c_i(t_1) - \langle R^c_i \rangle] / \sigma(R^c_i)$, where $\langle . \rangle $ is an average over a candidate period of time and $\sigma(.)$ is a standard deviation operator, computed over the same interval. Equity log returns are standardised analogously. Over any candidate smoothing window $S$, the cryptocurrency and equity correlation matrices, $\Omega^c$ and $\Omega^e$ are defined as follows:
\begin{align}
\label{eq:corrmatrix}
\Omega^c(t_1) = \frac{1}{S} \tilde{R}^c (t_1) \tilde{R}^{cT}(t_1), \forall t_1 \in S,...,T_1, \\
\Omega^e (t_2) = \frac{1}{S} \tilde{R}^e (t_2) \tilde{R}^{eT} (t_2), \forall t_2 \in S,...,T_2.
\end{align}

In our proceeding experiments, we set $S=150$, and study the evolution in structure of the cryptocurrency and equity correlation matrix time series. Previous literature has outlined potential sensitivity in findings to the parameter $S$ \cite{Fenn2011}. If $S$ is too small, although correlation structure can be computed dynamically, results may be noisy. Alternatively if $S$ is too large, our time-evolving model may be insensitive to abrupt changes in behaviours (such as the COVID-19 market crisis). We also wish to ensure that $Q = S/N \geq 1$ in subsequent experiments. 

We apply principal components analysis (PCA) to our two time-varying correlation matrices. In particular, we wish to study the evolution in the magnitude and explanatory variance of the correlation matrix's first eigenvalue. At every point in time each correlation matrix generates a sequence of ordered eigenvalues $\lambda^c_1(t_1),...,\lambda^c_N(t_1)$ and $\lambda^e_1(t_2),...,\lambda^e_K(t_2)$. At each point in time we normalize the eigenvalues in each collection by defining $\tilde{\lambda^c}_i (t_1) = \frac{\lambda^c_i (t_1) }{\sum^{N}_{z=1} \lambda^c_z (t_1) }$ and $\tilde{\lambda^e}_j (t_2) = \frac{\lambda^e_j (t_2) }{\sum^{K}_{z=1} \lambda^e_z (t_2) }$.

We now wish to compare the eigenspectrum and properties of our equity and cryptocurrency correlation matrices with that of random time series. Before proceeding, we provide some background on this problem from the literature. As pointed out in \cite{Kwapie2012}, decoupling correlation matrix behaviours into actual dependencies among the degrees of freedom and random correlation structure is a problem of significant difficulty. Many authors have employed \emph{random matrix theory} to address this problem \cite{Brody1981, Guhr1998}. The Marchenko-Pastur law is widely applied in the econophysics literature. We provide a basic overview of the law, following \cite{Bryson2021}, however more detailed exposition can be found in the original paper \cite{Marenko1967}. For any random matrix $X$ of dimension $p \times m$ where matrix elements are randomly samples from a standard normal distribution, the limiting distribution of the eigenspectrum, which we denote $\lambda_j(W)$, of the sample covariance matrix $W = \frac{1}{m}XX^{T}$ is determined by the Marchenko-Pastur law. This result is shown to be valid where the dimensions of our matrix $X$ increase to infinity, but the aspect ratio (the ratio of an objects size in different dimensions) remains constant. That is, as $p \rightarrow \infty$ and $p/m \rightarrow \lambda \in (0,\infty)$. From here, one can state with absolute certainty that the empirical spectral distribution of the $p \times p$ matrix $W$ will converge to a deterministic distribution, referred to as the Marchenko-Pastur law, with a single parameter $\lambda$. In the case $\lambda$ lies in $(0,1)$ the following will hold for all $x \in \mathbb{R}$:
\begin{align}
    F^{W}(x) := \frac{1}{p} \# \{1 \leq j \leq p : \lambda_j(W) \leq x \} \rightarrow \int^{x}_{\infty} f_{\lambda} (t) \,dt\
\end{align}
where $f_{\lambda}$ is the Marchenko-Pastur density
\begin{align}
    f_{\lambda} (x) = \frac{1}{2\pi \lambda x} \sqrt{[(\lambda_{+} - x)(x - \lambda_{-})]_{+}},
\end{align}
where $\lambda_{\pm} = (1 \pm \sqrt{\lambda})^2$.
We now turn to our application. The resulting correlation matrix among a collection of $N$ uncorrelated time series each of length $S$ where elements are drawn randomly from a Gaussian distribution, belongs to a class referred to as \emph{Wishart matrices} \cite{WISHART1928}. Assuming the respective constraints $Q^c=S/N \geq 1$ and $Q^e=S/K \geq 1$ are satisfied for both collections, in the limit where $N,K \to \infty$ and $S \to \infty$, the probability density functions of our two eigenvalue collections are given by:
\begin{align}
\label{eq:RMT_eigenvalues}
p(\lambda^c (t_1)) = \frac{Q^c}{2\pi \sigma^2 (\tilde{R}^c(t_1))} \frac{\sqrt{(\lambda^c_{+}(t_1) - \lambda^c(t_1))(\lambda^c_{-}(t_1) - \lambda^c(t_1))}}{\lambda^c(t_1)} \\
p(\lambda^e (t_2)) = \frac{Q^e}{2\pi \sigma^2 (\tilde{R}^e(t_2))} \frac{\sqrt{(\lambda^e_{+}(t_2) - \lambda^e(t_2))(\lambda^e_{-}(t_2) - \lambda^e(t_2))}}{\lambda^e(t_2)}. 
\end{align}

where $\sigma^2 (\tilde{R}^c)$ and $\sigma^2 (\tilde{R}^e)$ refer to the variance among the elements of $\tilde{R}^c$ and $\tilde{R}^e$ at each point in time $t_1$ and $t_2$. $\lambda_{+}$ and $\lambda_{-}$ refer to the maximum and minimum eigenvalues of the respective correlation matrices, and are computed as follows:
\begin{align}
\label{eq:RMT_max_eigenvalues}
\lambda^c_{\pm} (t_1) = \sigma^{2}(\tilde{R}^c(t_1)) \bigg( 1 + \frac{1}{Q^c} \pm 2 \sqrt{\frac{1}{Q^c}}\bigg),   \\
\lambda^e_{\pm} (t_2) = \sigma^{2}(\tilde{R}^e(t_2)) \bigg( 1 + \frac{1}{Q^e} \pm 2 \sqrt{\frac{1}{Q^e}}\bigg). 
\end{align}

As we apply our random matrix analysis in a rolling capacity, we must ensure that $Q^c$ and $Q^e$ are greater than 1 with $S=150$. Given that our cryptocurrency collection consists of $N=45$ unique cryptocurrencies and our equity collection consists of $K=77$ unique equities, $Q^c = 3.3$ and $Q^e = 1.95$. It is well understood that when $Q=1$, the minimum eigenvalue is $\lambda_{-} = 0$ and the maximum eigenvalue is $\lambda_{+} = 4 \sigma^2(\tilde{R})$. For the two collections of securities we study, $\lambda^c_{+} = 1.45$ and $\lambda^e_{+} = 1.75$.

\begin{figure}
    \centering
    \begin{subfigure}[b]{0.75\textwidth}
        \includegraphics[width=\textwidth]{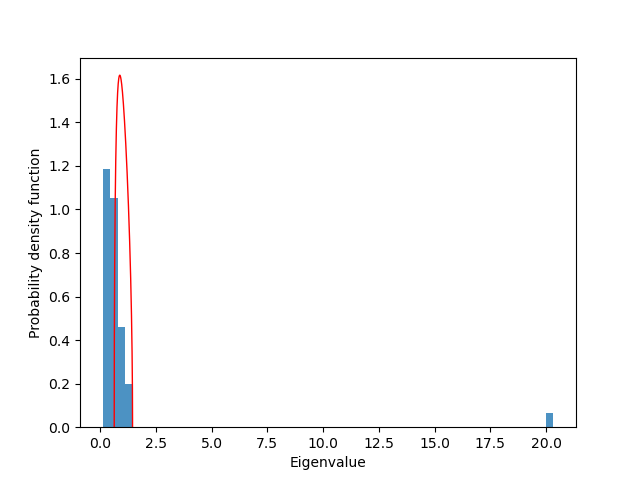}
        \caption{}
        \label{fig:Eigenvalue_distribution_crypto}
    \end{subfigure}
    \begin{subfigure}[b]{0.75\textwidth}
        \includegraphics[width=\textwidth]{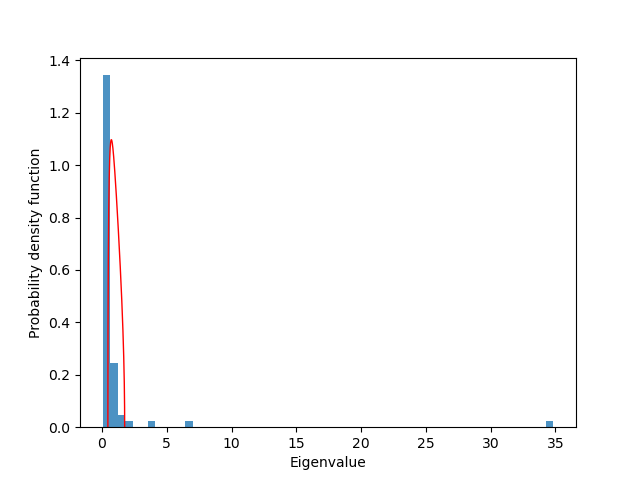}
        \caption{}
        \label{fig:Eigenvalue_distribution_equity}
    \end{subfigure}
    \caption{Plot of theoretical density and empirical density for (a) cryptocurrencies and (b) equities between 01-01-2019 to 01-12-2021. The equity collection clearly exhibits a greater number of non-random eigenvalues.}
    \label{fig:Eigenvalue_distribution_stationary}
\end{figure}

Figure \ref{fig:Eigenvalue_distribution_stationary} shows the theoretical and empirical probability density functions of each collections' eigenvalues. Several noteable differences between cryptocurrency and equity behaviours are observed. First, the cryptocurrency eigenvalues, which are shown in Figure \ref{fig:Eigenvalue_distribution_crypto}, exhibit one dominant eigenvalue, $\lambda^c_1$, which is significantly greater than the theoretical upper bound, $\lambda_{+}^c$. This confirms the results demonstrated by \cite{Drod2020} in Figure 3 of their work. By contrast, the equity collection (shown in Figure \ref{fig:Eigenvalue_distribution_equity}) consists of four eigenvalues ($\lambda_1^e, \lambda_2^e, \lambda_3^e$ and $\lambda_4^e$) all of which are larger in magnitude than the theoretical largest eigenvalue, $\lambda^e_{+}$. Our finding for the equity market is consistent with the results shown in \cite{Kwapie2012}. The authors highlight that the presence of several significant eigenvalues (beyond the standard Wishart regime) corresponds to the effect of grouping stocks within various market sectors. Further discussion on this phenomenon, and similar phenomena, can be found in their extensive review. There is also significant literature detailing more complex models surrounding Wishart matrices \cite{Recher2010, Hachem2016, Wirtz2013, Waltner2015}. The difference in the number of non-random eigenvalues between our collection is striking, and may be suggestive of a more profound difference in the impact of exogenous variables on both collections. Given that the first (dominant) eigenvalue represents the collective behaviour of the market \cite{Fenn2011}, the existence of meaningful subsequent eigenvalues in equities may highlight the sophistication of the market (particularly in comparison to cryptocurrencies), where non-market related collective behaviours influence the correlation structure between underlying securities. By contrast, the cryptocurrency market's exhibiting one non-random eigenvalue draws attention to the overwhelming importance of the `market' factor among the collection.


\begin{figure}
    \centering
    \begin{subfigure}[b]{0.75\textwidth}
        \includegraphics[width=\textwidth]{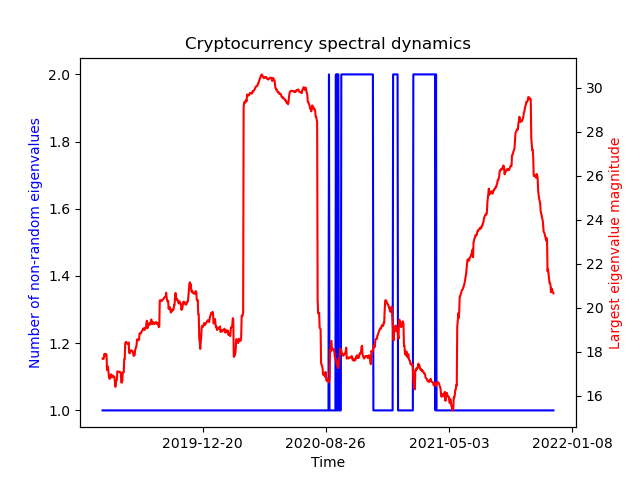}
        \caption{}
        \label{fig:Cryptocurrency_time_varying_RMT}
    \end{subfigure}
    \begin{subfigure}[b]{0.75\textwidth}
        \includegraphics[width=\textwidth]{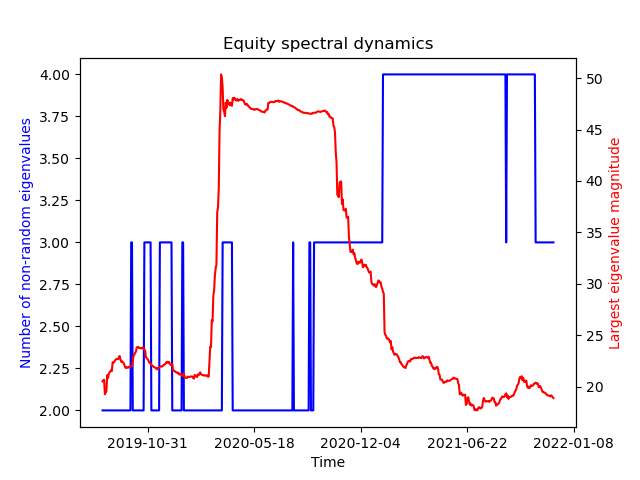}
        \caption{}
        \label{fig:Equity_time_varying_RMT}
    \end{subfigure}
    \caption{Time-varying computation of leading eigenvalue magnitude and number of non-random eigenvalues for (a) cryptocurrencies and (b) equities. The equity collection consistently produces a larger number of non-random eigenvalues.}
    \label{fig:RMT_time_varying}
\end{figure}

Having studied the distribution of correlation matrix eigenvalues in the stationary case, we now investigate how this structure evolves over time in both collections. Figure \ref{fig:RMT_time_varying} displays the evolution in the magnitude of the dominant eigenvalue ($\lambda_1^c(t_1)$ and $\lambda_1^e(t_2)$) and the number of eigenvalues greater than the theoretical maximum eigenvalue $\lambda^c_{+}(t_1)$ and $\lambda^e_{+}(t_2)$, respectively. There are several significant findings. The number of non-random eigenvalues in the cryptocurrency collection, shown in Figure \ref{fig:Cryptocurrency_time_varying_RMT}, ranges between 1 and 2. By contrast, the equity collection exhibits between 2 and 4 non-random eigenvalues over the entirety of our analysis window. This is shown in Figure \ref{fig:Equity_time_varying_RMT}. This confirms the persistence of the findings first highlighted in our static analysis, with the equity collection consistently exhibiting a larger number of non-random eigenvalues than the cryptocurrency collection. Both figures reveal a consistently inverse relationship between the number of non-random eigenvalues and the magnitude of the first eigenvalue. The intuition behind this relationship is interesting and open to a variety of interpretations. One possible argument is that when there is an increase in the magnitude of the dominant eigenvalue, the collective strength of the market increases. Given that this phenomenon is shown to coincide with a smaller number of non-random eigenvalues, one could assume that when the collective strength of the market is high - there is less opportunity for alternative factors to influence correlation behaviours of underlying securities. Given that spikes in the first eigenvalue tend to coincide with bear markets and the indiscriminate selling of assets, one can see that during market crises, exogenous factors other than the collective strength of the market (represented by eigenvalues $\lambda_2$,...,$\lambda_m$ such that $\lambda_m > \lambda_{+}$) have less influence on correlation dynamics. This pattern appears to exist in both the cryptocurrency and equity markets. 


\section{Sector evolutionary dynamics}
\label{Evolutionary_dynamics_sector}

In this section we wish to explore the potential heterogeneity of each time series collection, and compare the evolution in the strength of market dynamics, represented by the correlation matrix's dominant eigenvalue. We begin by partitioning our collection of cryptocurrencies into 8 sectors, and computing a daily rolling correlation matrix for each sector, again setting $S=150$. Let our cryptocurrency correlation matrices at time $t_1$ be $\Omega^{c,\psi_l} (t_1)$, where $\psi_{l}: l=1,...,8$ refers to any one of $L=8$ potential cryptocurrency sectors. Similarly, let $\Omega^{e,\psi_p} (t_2)$ refer to one of $\psi_{p}: p=1,...,11$ be any one of $P=11$ equity sector correlation matrices at time $t_2$. We are particularly interested in studying the evolution of collective behaviour within candidate sectors, and how much this evolution varies between different sectors. Let $\tilde{\lambda}^{c,\psi_l}_1 (t_1)$ and $\tilde{\lambda}^{e,\psi_p}_1 (t_2)$ be the explanatory variance exhibited by the first eigenvalue of each sector's correlation matrix at any point in time. This normalization is computed as described in Section \ref{RMT}. We determine the affinity between sectors' evolutionary collective similarity by computing an $L^1$ norm between any two paths of the first eigenvalue's explanatory variance. Our two distance matrices are computed as follows: 

\begin{figure}
    \centering
    \begin{subfigure}[b]{0.75\textwidth}
        \includegraphics[width=\textwidth]{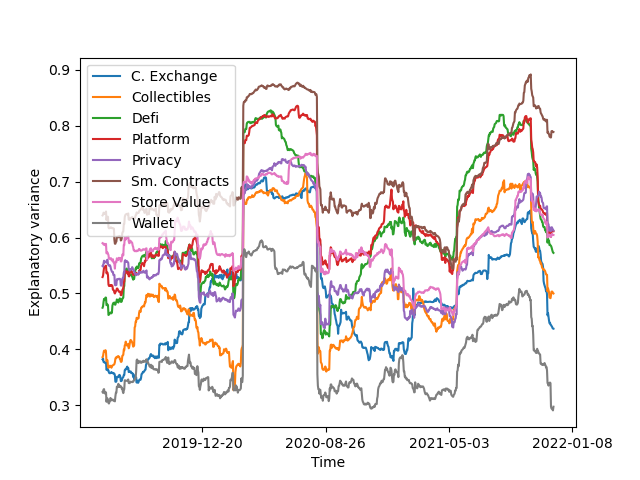}
        \caption{}
        \label{fig:Explanatory_variance_crypto}
    \end{subfigure}
    \begin{subfigure}[b]{0.75\textwidth}
        \includegraphics[width=\textwidth]{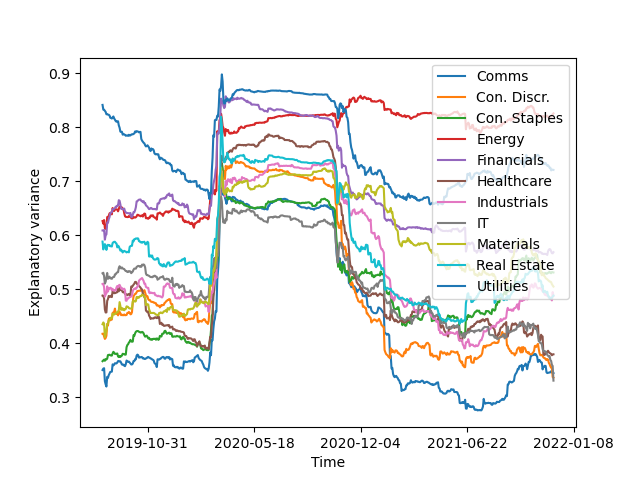}
        \caption{}
        \label{fig:Explanatory_variance_equity}
    \end{subfigure}
    \caption{Time-varying explanatory variance exhibited by the first eigenvalue of each sector's correlation matrix among (a) cryptocurrencies and (b) equities. The cryptocurrency market is partitioned into 8 sub-groups (sectors) including: centralized exchange, collectibles/NFTs, Decentralized finance, Platforms, Privacy, Smart contracts, Store of value and Wallet. Equities are classified based on their GICS classification: Communication services, Consumer discretionary, Consumer staples, Energy, Financials, Healthcare, Industrials, Information technology, Materials, Real estate and Utilities.}
    \label{fig:Explanatory_variance_sectors}
\end{figure}

\begin{align}
\label{eq:DistanceMatrix}
D_{ij}^{c, \tilde{\lambda_1}} = \frac{1}{T_1 - S} \sum^{T_1}_{t_1=S} \bigg| \tilde{\lambda}^{c,\psi_i}_1 (t_1) - \tilde{\lambda}^{c,\psi_j}_1 (t_1) \bigg|, \\
D_{ij}^{e, \tilde{\lambda_1}} = \frac{1}{T_2 - S} \sum^{T_2}_{t_2=S} \bigg| \tilde{\lambda}^{e,\psi_i}_1 (t_2) - \tilde{\lambda}^{e,\psi_j}_1 (t_2) \bigg|.
\end{align}

\begin{figure}
    \centering
    \begin{subfigure}[b]{0.75\textwidth}
        \includegraphics[width=\textwidth]{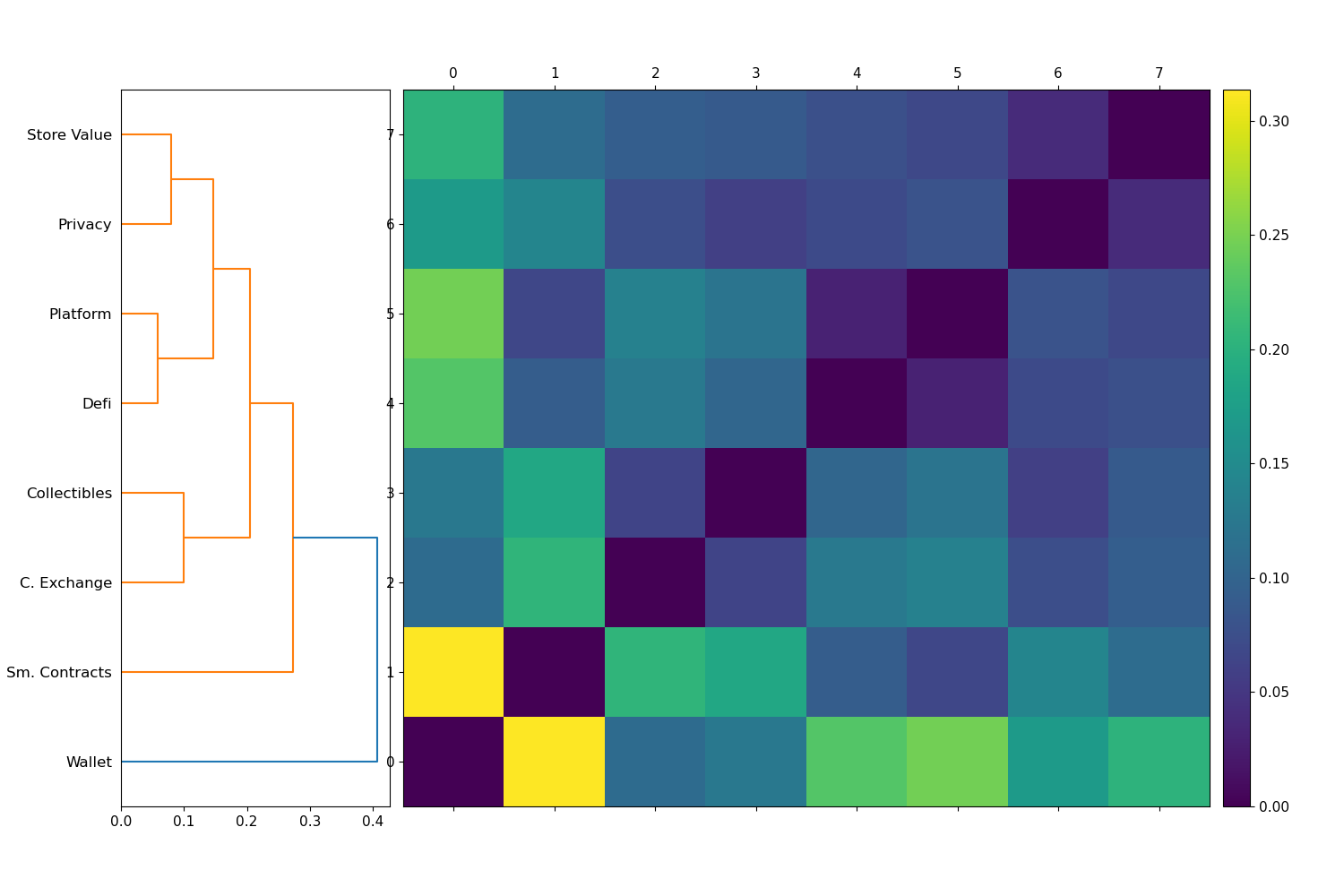}
        \caption{}
        \label{fig:lambda_1_crypto_dendrogram}
    \end{subfigure}
    \begin{subfigure}[b]{0.75\textwidth}
        \includegraphics[width=\textwidth]{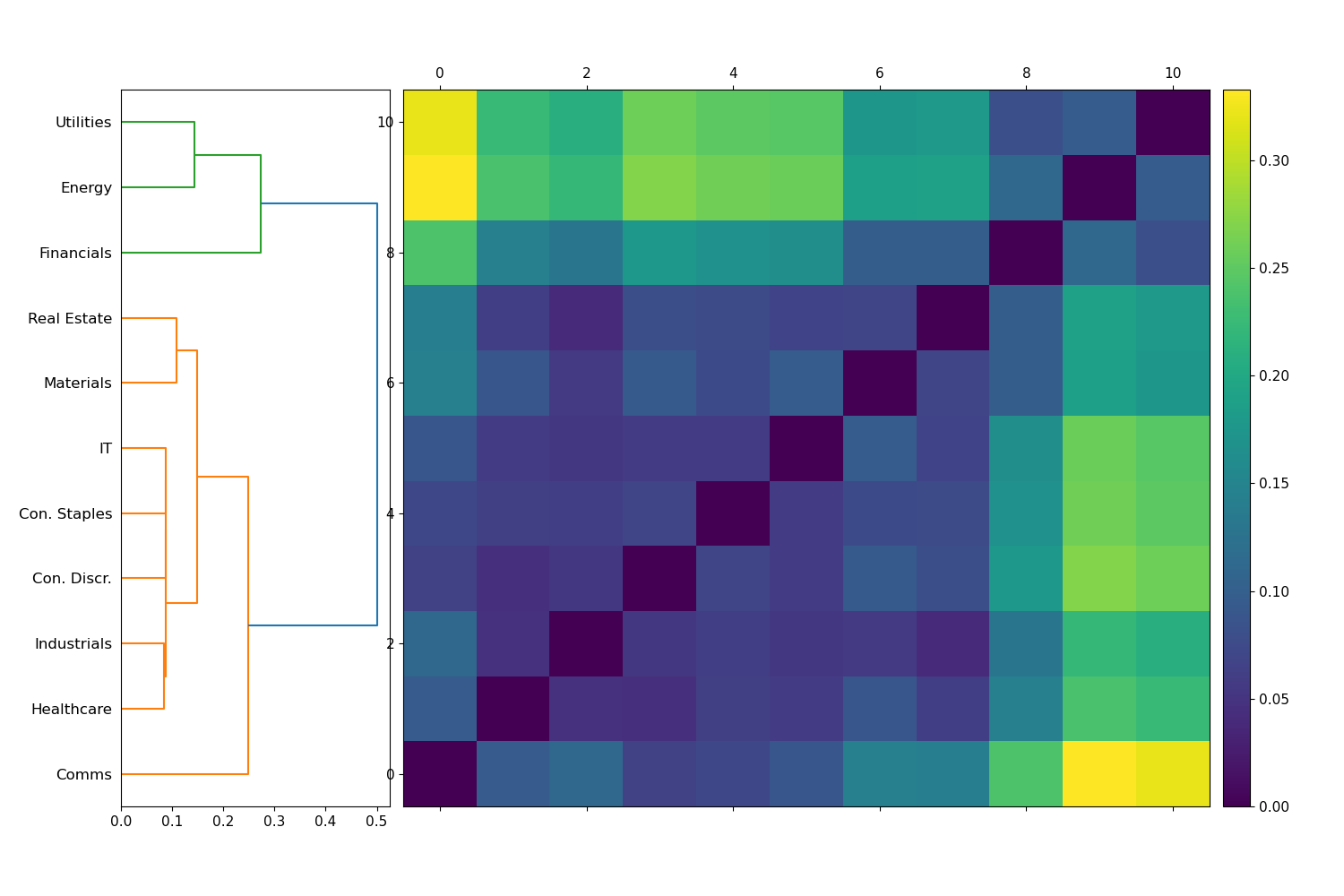}
        \caption{}
        \label{fig:lambda_1_equity_dendrogram}
    \end{subfigure}
    \caption{Hierarchical clustering applied to our two distance matrices (a) $D^{c, \tilde{\lambda_1}}$ and (b) $D^{e, \tilde{\lambda_1}}$. The dendrograms reveal some notable differences in the heterogeneity of evolutionary dynamics of cryptocurrency and equity sectors. The cryptocurrency collection exhibits a propensity for coupling behaviour between sectors, while the equity collection produces a predominant cluster of high affinity (and a small collection of outlier sectors).}
    \label{fig:lambda_1_dendrograms}
\end{figure}

Figure \ref{fig:Explanatory_variance_sectors} shows the time-varying explanatory variance exhibited by respective cryptocurrency and equity sectors. There are some broad similarities in the evolution of respective sectors and the time-varying affinity among both collections. Both the cryptocurrency and equity collections, shown in Figures \ref{fig:Explanatory_variance_crypto} and \ref{fig:Explanatory_variance_equity}, display surprising consistency in the ranking of each sector's first eigenvalue explanatory variance over time. 
In the cryptocurrency sector, smart contract-themed cryptocurrencies consistently display the highest collective market strength, while wallet-themed cryptocurrencies consistently display the lowest. Within our equity collection, utilities display the strongest collective market dynamics, while our sample of communications stocks exhibits the weakest collective market strength. Both collections produce a similar range in $\tilde{\lambda}_1$ scores over time, suggesting that each asset class would present similarly difficult problems for portfolio managers seeking diversification among security returns. One notable difference between the collections however, is the range in $\tilde{\lambda}_1$ values during the COVID-19 market crisis. The range among cryptocurrency sectors is approximately 0.3, while the range among respective equity sectors is approximately 0.2. This may suggest that for portfolio managers focused solely on individual asset classes, cryptocurrencies could provide more opportunity for diversifying across collective behaviours during market crises. 


We further investigate the similarity in evolutionary market dynamics by applying hierarchical clustering to our two distance matrices, $D^{c, \tilde{\lambda_1}}$ and $D^{e, \tilde{\lambda_1}}$. Figures \ref{fig:lambda_1_crypto_dendrogram} and \ref{fig:lambda_1_equity_dendrogram} show the cryptocurrency and equity evolutionary collective behaviour dendrograms respectively. Both figures displays broad similarity in the collective affinity between respective sectors, with the scale of both dendrograms ranging from 0 to 0.32 and 0 to 0.35 respectively. Figure \ref{fig:lambda_1_crypto_dendrogram} reveals the propensity for coupling among cryptocurrency sectors. Three notable pairs are revealed between Store of value and privacy, platform and decentralized finance and collectibles/nfts and centralized exchange sectors. The smart contracts and wallet-themed cryptocurrencies exhibit more anomalous evolutionary collective dynamics with respect to the rest of the collection. By contrast the collection of equities, which are clustered together in Figure \ref{fig:lambda_1_crypto_dendrogram}, displays a significantly different cluster structure. There is a predominant cluster of similarity among 8 sectors (real estate, materials, information technology, consumer staples, consumer discretionary, industrials, healthcare and communications), with a small collection of oultier sectors which includes utilities, energy and financials. The predominant similarity between constituent sectors of the primary cluster are consistent with the $\tilde{\lambda^e_1}$ paths shown in Figure \ref{fig:Explanatory_variance_equity}, where collective behaviours are shown to be of closer proximity than that of the cryptocurrency market. 


\section{Structural breaks and spectral regime switching}
\label{Structural_breaks_non_stationarity}

\subsection{Structural break dynamics}
In this section we investigate the structural break dynamics and regime switching behaviour of cryptocurrencies and equities. We extend a recently introduced framework \cite{James_mj_wasserstein_2022}, which marries a Bayesian change point detection algorithm \cite{james2021_spectral, Rosen2012} with a new distance metric which quantifies distance between sets of probability distributions. The Bayesian changepoint detection framework utilizes a reversible jump Markov chain Monte Carlo (RJMCMC) algorithm to produce a distribution over possible models, where each model consists of a varying number of change points, $m$. In all RJMCMC experiments we run 10,000 iterations and discard the first 5,000 of these iterations (burn-in). We select the \emph{maximum a posteriori} model for each time series, and accordingly determine an appropriate number of change points for each cryptocurrency and equity security respectively. In subsequent experiments we elect one representative cryptocurrency/equity from each sector and apply our respective methodology. This is primarily for computational savings, given the high cost in running the Reversible Jump Markov Chain Monte Carlo algorithm. 

\begin{figure}
    \centering
    \begin{subfigure}[b]{0.48\textwidth}
        \includegraphics[width=\textwidth]{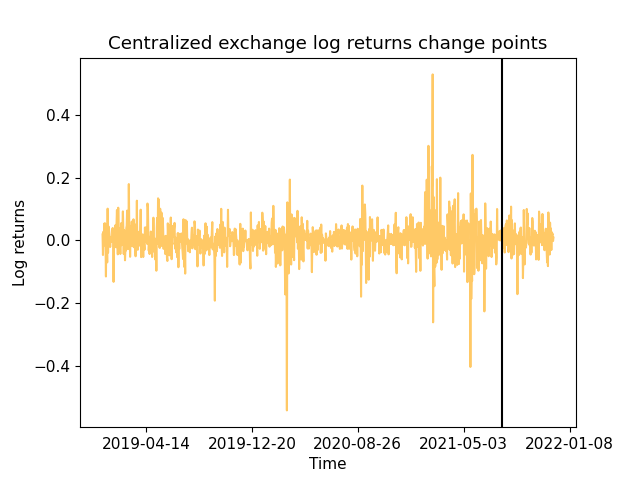}
        \caption{}
        \label{fig:CP_exchange}
    \end{subfigure}
    \begin{subfigure}[b]{0.48\textwidth}
        \includegraphics[width=\textwidth]{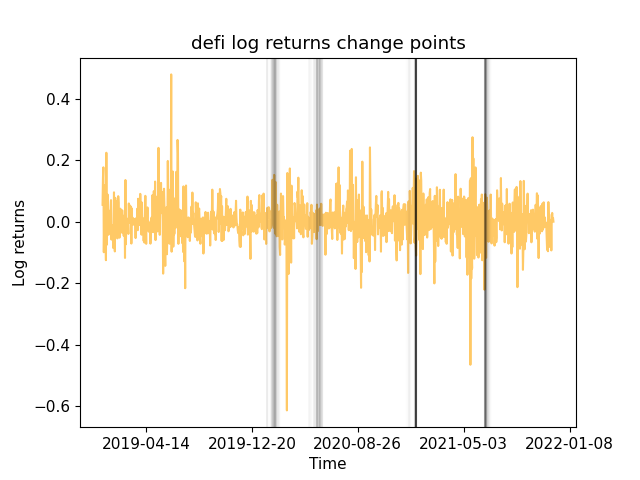}
        \caption{}
        \label{fig:CP_defi}
    \end{subfigure}
    \begin{subfigure}[b]{0.48\textwidth}
        \includegraphics[width=\textwidth]{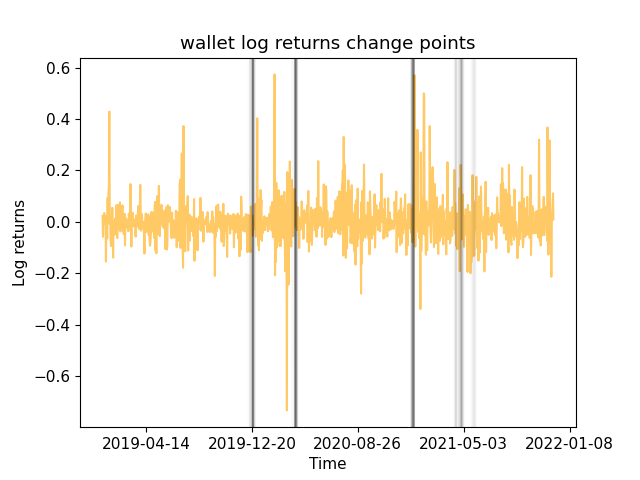}
        \caption{}
        \label{fig:CP_wallet}
    \end{subfigure}
    \begin{subfigure}[b]{0.48\textwidth}
        \includegraphics[width=\textwidth]{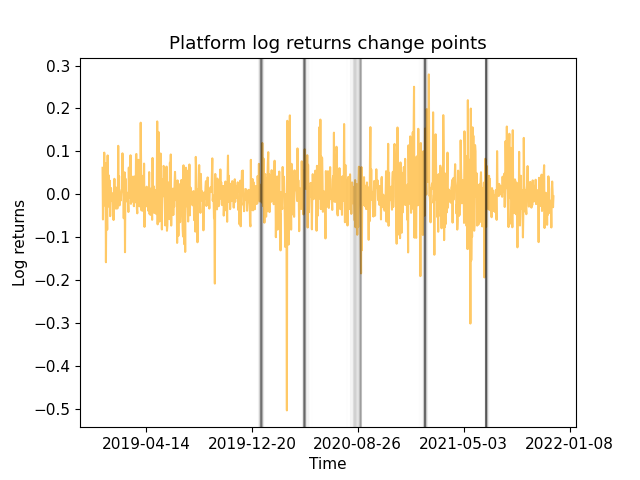}
        \caption{}
        \label{fig:CP_platform}
    \end{subfigure}
    \caption{A sample of cryptocurrency sector log returns with annotated change points (a) Centralized exchange (b) Decentralized finance, (c) Wallet, (d) Platform. The transparency of the black vertical lines indicate the probability of a changepoint at any candidate location (darker colours are consistent with greater probability mass). Although the four sectors above all display structural breaks corresponding to the beginning and end of the COVID-19 market crash, there is significantly greater variability in the number and location of structural breaks than that exhibited by the equity collection.}
    \label{fig:Crypto_structural_breaks}
\end{figure}

\begin{figure}
    \centering
    \begin{subfigure}[b]{0.48\textwidth}
        \includegraphics[width=\textwidth]{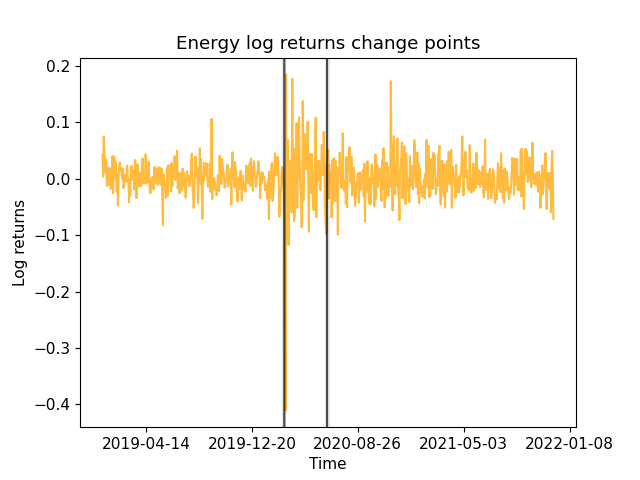}
        \caption{}
        \label{fig:CP_Energy}
    \end{subfigure}
    \begin{subfigure}[b]{0.48\textwidth}
        \includegraphics[width=\textwidth]{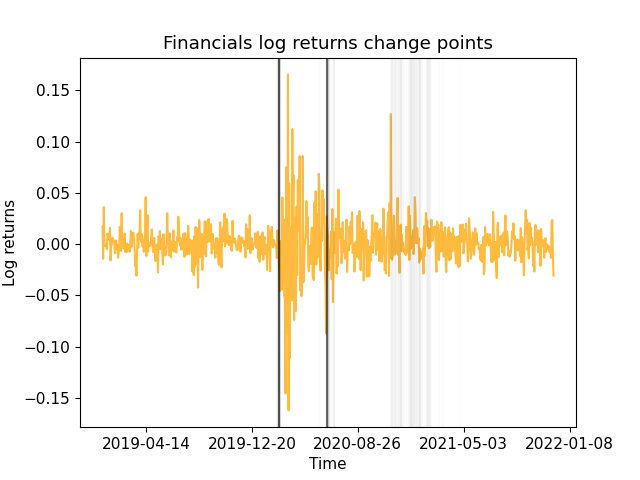}
        \caption{}
        \label{fig:CP_Financials}
    \end{subfigure}
    \begin{subfigure}[b]{0.48\textwidth}
        \includegraphics[width=\textwidth]{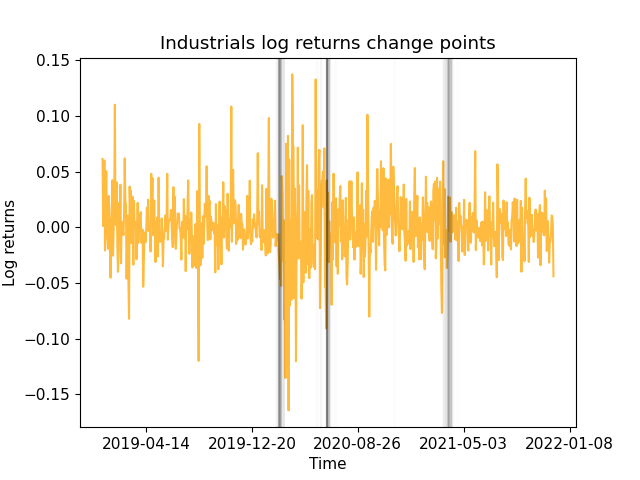}
        \caption{}
        \label{fig:CP_Industrials}
    \end{subfigure}
    \begin{subfigure}[b]{0.48\textwidth}
        \includegraphics[width=\textwidth]{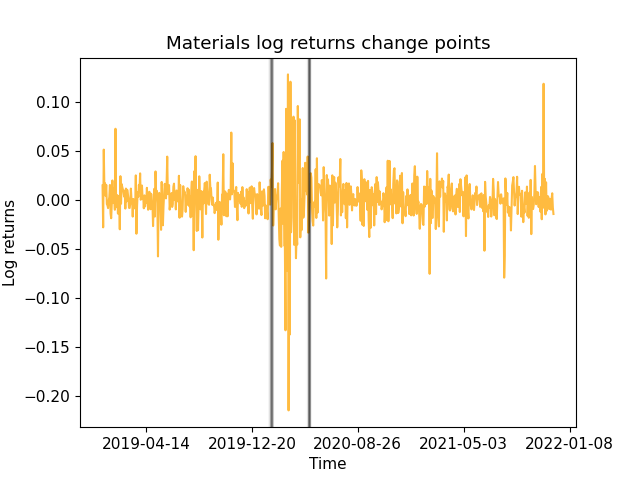}
        \caption{}
        \label{fig:CP_Materials}
    \end{subfigure}
    \caption{A sample of equity sector log returns with annotated change points (a) Energy (b) Financials, (c) Industrials, (d) Materials. The transparency of the black vertical lines indicate the probability of a changepoint at any candidate location (darker colours are consistent with greater probability mass). All equity sectors exhibit two clear structural breaks indicating the beginning and end of the COVID-19 market crash.}
    \label{fig:Equity_structural_breaks}
\end{figure}

The location of each changepoint is associated with a probability density function. Let $\tilde{S}, \tilde{T}$ refer to any two sets with uncertainty within our respective collections. For any candidate time series pair, let $\tilde{S} = \{f_1,...,f_a\}$ and $\tilde{T} = \{g_1,...,g_b\}$, be our two sets with uncertainty which contain $a$ and $b$ probability distributions respectively. Within each set, the probability distributions $1,...,a$ and $1,...,b$ associate a collection of time series indices with a changepoint probability. In the proceeding experiments, we set the polynomial order $o=1$, and compute two MJ-Wasserstein distance matrices as follows:

\begin{align}
\label{eq:MJW_DistanceMatrix}
D^{c,MJW} (\tilde{S}^c_i, \tilde{T}^c_j) &= \Bigg(\frac{\sum_{g\in \tilde{T}^c_j} d_W(g,\tilde{S}^c_i)^o}{2|\tilde{T}^c_j|} + \frac{\sum_{{f} \in \tilde{S}^c_i} d_W(f,\tilde{T}^c_j)^o}{2|\tilde{S}^c_i|} \Bigg)^{\frac{1}{o}} \\
D^{e,MJW} (\tilde{S}^e_i, \tilde{T}^e_j) &= \Bigg(\frac{\sum_{g\in \tilde{T}^e_j} d_W(g,\tilde{S}^e_i)^o}{2|\tilde{T}^e_j|} + \frac{\sum_{{f} \in \tilde{S}^e_i} d_W(f,\tilde{T}^e_j)^o}{2|\tilde{S}^e_i|} \Bigg)^{\frac{1}{o}}. 
\end{align}

In the two equations above (outlining the distance among our cryptocurrency and equity collections), for each distribution $f_i$ in $\tilde{S}$, the nearest distribution $g_j$ in $\tilde{T}$ is identified according to the Wasserstein distance metric, and the minimal distance between any two sets is computed. This allows us to determine the similarity between any two time series' collection of changepoints. Importantly, this metric allows us to capture the uncertainty associated with the location of each change point within any time series' changepoint profile. 

\begin{table}
\centering
\begin{tabular}{ |p{3.5cm}||p{3.8cm}|p{2cm}|}
 \hline
 \multicolumn{3}{|c|}{Estimated number of regimes by asset class sector} \\
 \hline
 Asset class & Sector & \# Regimes  \\
 \hline
 Cryptocurrency & Centralized exchange & 6 \\
 Cryptocurrency & Collectibles/NFTs & 6 \\
 Cryptocurrency & Decentralized finance & 2 \\
 Cryptocurrency & Platform & 3 \\
 Cryptocurrency & Privacy & 5 \\
 Cryptocurrency & Smart contracts & 5 \\
 Cryptocurrency & Store of value & 2 \\
 Cryptocurrency & Wallet & 5 \\
 Equity & Communication services & 4 \\
 Equity & Consumer discretionary & 3 \\
 Equity & Consumer staples & 4 \\
 Equity & Energy & 3 \\
 Equity & Financials & 4 \\
 Equity & Healthcare & 4 \\
 Equity & Industrials & 4 \\
 Equity & Information Technology & 4 \\
 Equity & Materials & 3 \\
 Equity & Real estate & 4 \\
 Equity & Utilities & 4 \\
\hline
\end{tabular}
\caption{Estimated number of sector regimes based on Reversible Jump MCMC algorithm's determination.}
\label{tab:Regimes_table}
\end{table}

The methodology we use highlights marked differences between cryptocurrency and equity sectors' structural breaks. Figure \ref{fig:Crypto_structural_breaks} shows the log returns and estimated structural breaks of four cryptocurrency sectors; centralized exchange (Figure \ref{fig:CP_exchange}), decentralized finance (Figure \ref{fig:CP_defi}), wallet (Figure \ref{fig:CP_wallet}) and platform (Figure \ref{fig:CP_platform}). There is clearly great variability in the number, location and associated uncertainty of each sector's structural breaks. The Centralized exchange sector exhibits just one change point, while decentralized finance and wallet both exhibit four and platform exhibits five. Turning to equities, Figure \ref{fig:Equity_structural_breaks} shows four equity sectors' empirical log returns with annotated change points determined by the change point algorithm. It is clear from the Figure that energy (Figure \ref{fig:CP_Energy}), financials (Figure \ref{fig:CP_Financials}), industrials (Figure \ref{fig:CP_Industrials}) and materials (Figure \ref{fig:CP_Materials}) all display highly similar change point dynamics around the COVID-19 market crisis. The energy sector and materials sector both exhibit two structural breaks, while financials and industrials exhibit three change points. Although both the financials and materials sectors are determined to exhibit an additional (similarly located) structural break corresponding to a shift in these sectors' log returns behaviours, in general, equity sectors appear to exhibit greater affinity in their structural break profile than that of the cryptocurrency sectors. However, the financials and industrials sectors both exhibit one additional changepoint located in early 2021 - corresponding to a shift in the behaviours of these sectors' log returns. 

\begin{figure}
    \centering
    \begin{subfigure}[b]{0.75\textwidth}
        \includegraphics[width=\textwidth]{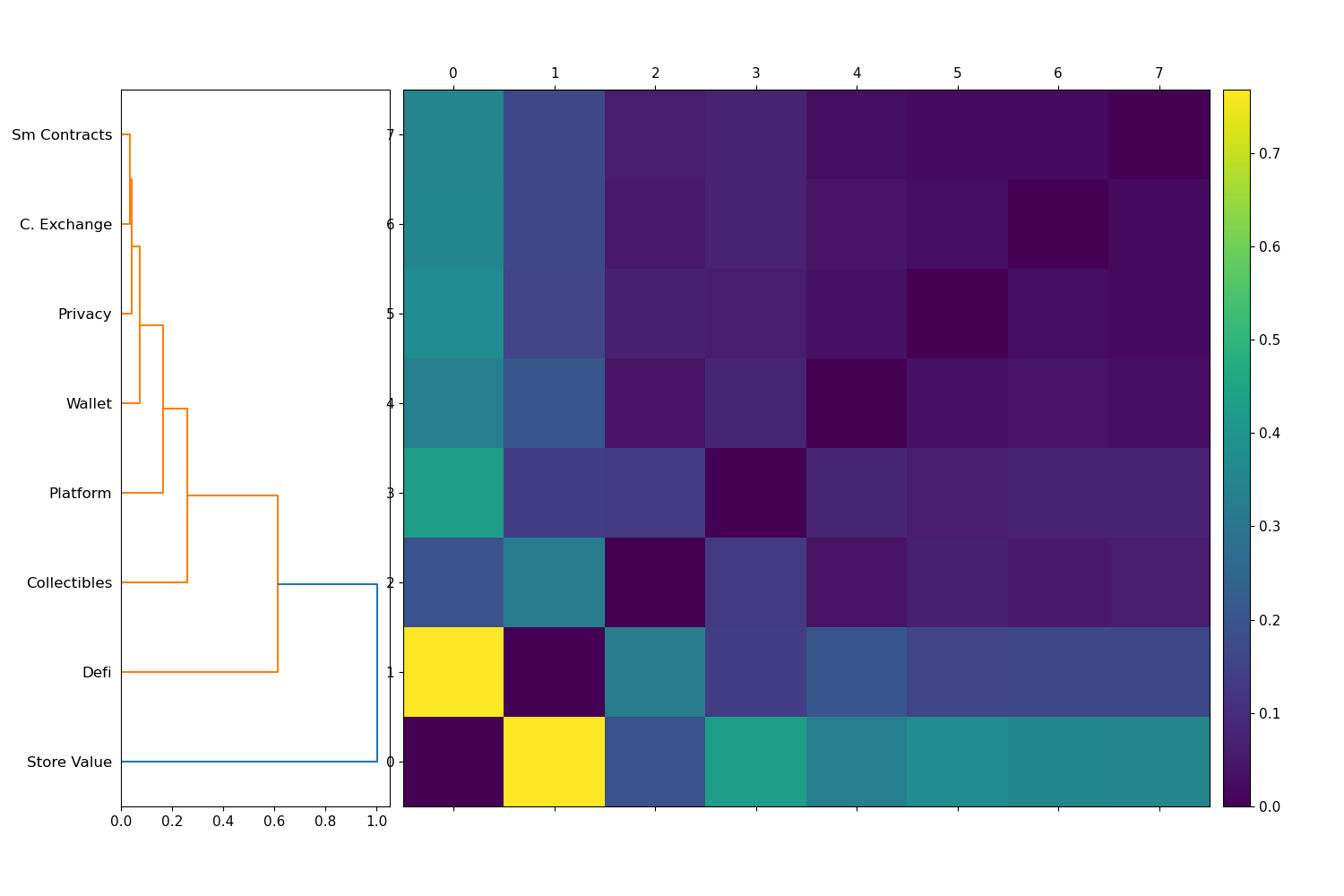}
        \caption{}
        \label{fig:Structural_breaks_crypto_mj}
    \end{subfigure}
    \begin{subfigure}[b]{0.75\textwidth}
        \includegraphics[width=\textwidth]{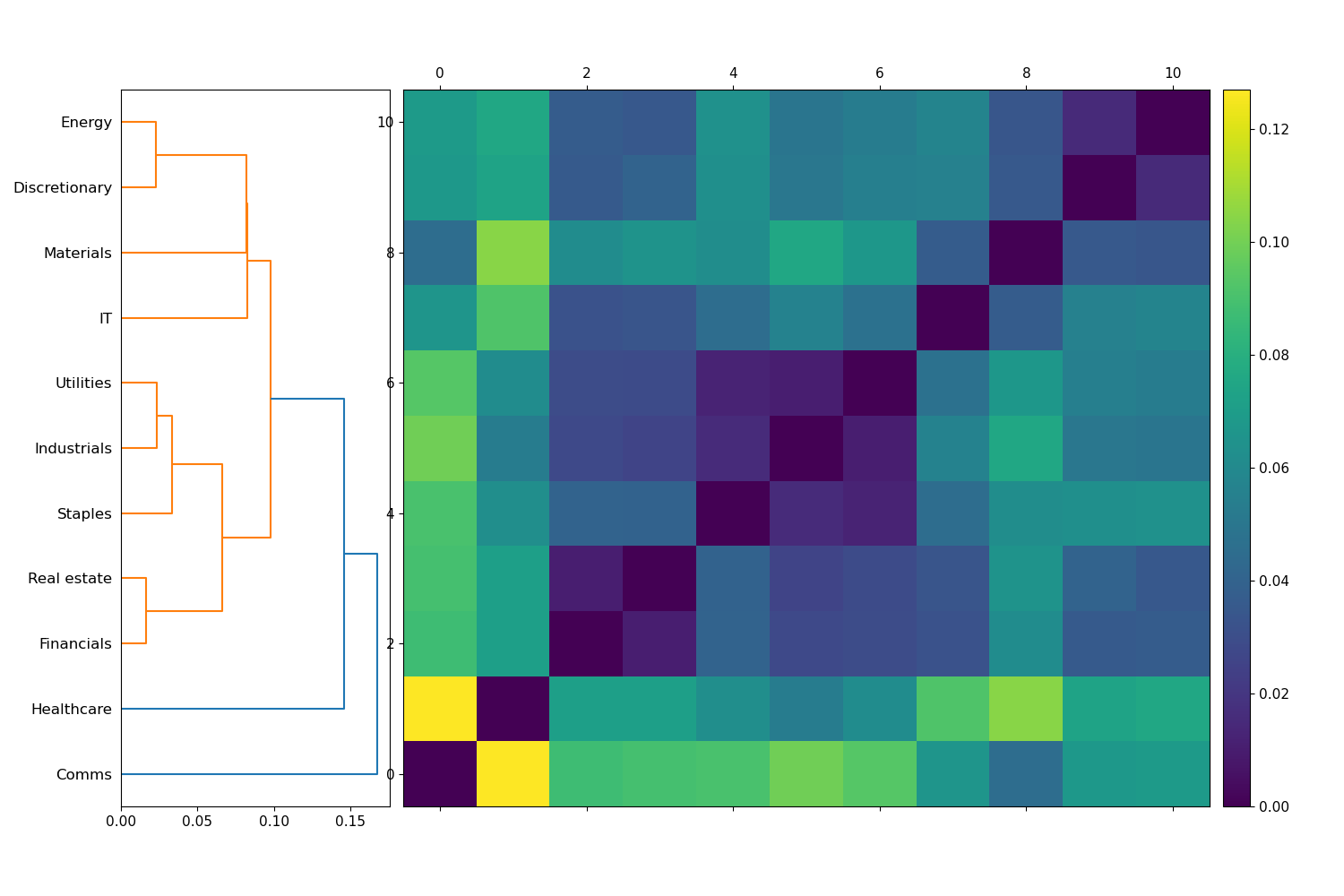}
        \caption{}
        \label{fig:Structural_breaks_equity_mj}
    \end{subfigure}
    \caption{Hierarchical clustering applied to cryptocurrency and equity structural break distance matrices, $D^{c, MJW}$ and $D^{e, MJW}$. The two dendrograms highlight significantly greater similarity between equity sector structural breaks than that exhibited by cryptocurrency sectors. }
    \label{fig:Dendrogram_mj_wasserstein}
\end{figure}

We further investigate differences in the propensity for various sectors' structural breaks in Table \ref{tab:Regimes_table}, which indicates the determined number of locally stationary segments. Among cryptocurrency sectors, the average number of regimes is 4.25 with a modal value of 5. The average number of equity sector regimes by contrast is 3.7, with a modal value of 4. The stark difference in structural break similarity is evidenced in Figure \ref{fig:Dendrogram_mj_wasserstein}, which highlights significantly greater similarity in the structural break profile of equities. Such analysis considers both the number and location of such change points. An indication of this, is the comparative scale of both dendrograms. The cryptocurrency structural break dendrogram, shown in Figure \ref{fig:Structural_breaks_crypto_mj}, has a range of 0.75, while the equity structural break dendrogram, shown in Figure \ref{fig:Structural_breaks_equity_mj}, has a range of just 0.13. One may conclude from this analysis, that cryptocurrency sectors are more prone to regime switching behaviours, and display less similarity in their structural break profile than equity sectors. 



\subsection{Spectral regime switching}
Next, we seek to further investigate the collective similarity in evolutionary power spectra among cryptocurrency and equity sectors. In the previous section, we focused solely on the measurement between sets of structural breaks, ignoring all the data within any candidate segment. In this section, we study the evolutionary \emph{power spectral density} of each sector and determine respective distances. This measure may provide a better indication of the similarity in the behaviours between any two sectors, where regime switching (or abrupt changes in behaviour) may exist.

Unlike many regime switching methods which rely on strong parametric assumptions, our nonparametric Bayesian framework determines the number of regimes associated with each time series. Structural breaks are determined based on changes in the time series' evolutionary power spectral density. Accordingly, we can study the similarity in the number of regimes between various sectors, and the similarity in their evolutionary power spectral density. Let the time-varying power spectrum for each cryptocurrency sector be $f^c_{\psi_l} (\nu, t_1)$ where the pair $(\nu, t_1)$ refers to a vector of respective frequency components at points in time $t_1$, and $\psi_l: l=1,...,8$ indexes one of $L=8$ possible cryptocurrency sectors. Similarly, let $f^e_{\psi_p}(\nu, t_2)$ refer to each equity sector's evolutionary power spectrum, where again $(\nu, t_2)$ refers to a vector of frequency components at various point in time $t_2$, and $\psi_p: p=1,...,11$ indexes one of $P=11$ possible equity sectors.

\begin{align}
\label{eq:SpectralDistanceMatrix}
D_{ij}^{c, ({\nu,t_1})} = \frac{1}{T_1 \times K} \sum^{T_1}_{t_1=1} \sum^{K}_{k=1} \bigg| f_{\psi_i}^c (\nu_k, t_1) - f_{\psi_j}^c (\nu_k, t_1) \bigg|, \\
D_{ij}^{e, ({\nu,t_2})} = \frac{1}{T_2 \times K} \sum^{T_2}_{t_2=1} \sum^{K}_{k=1} \bigg| f_{\psi_i}^e (\nu_k, t_2) - f_{\psi_j}^e (\nu_k, t_2) \bigg|. 
\end{align}

\begin{figure}
    \centering
    \begin{subfigure}[b]{0.75\textwidth}
        \includegraphics[width=\textwidth]{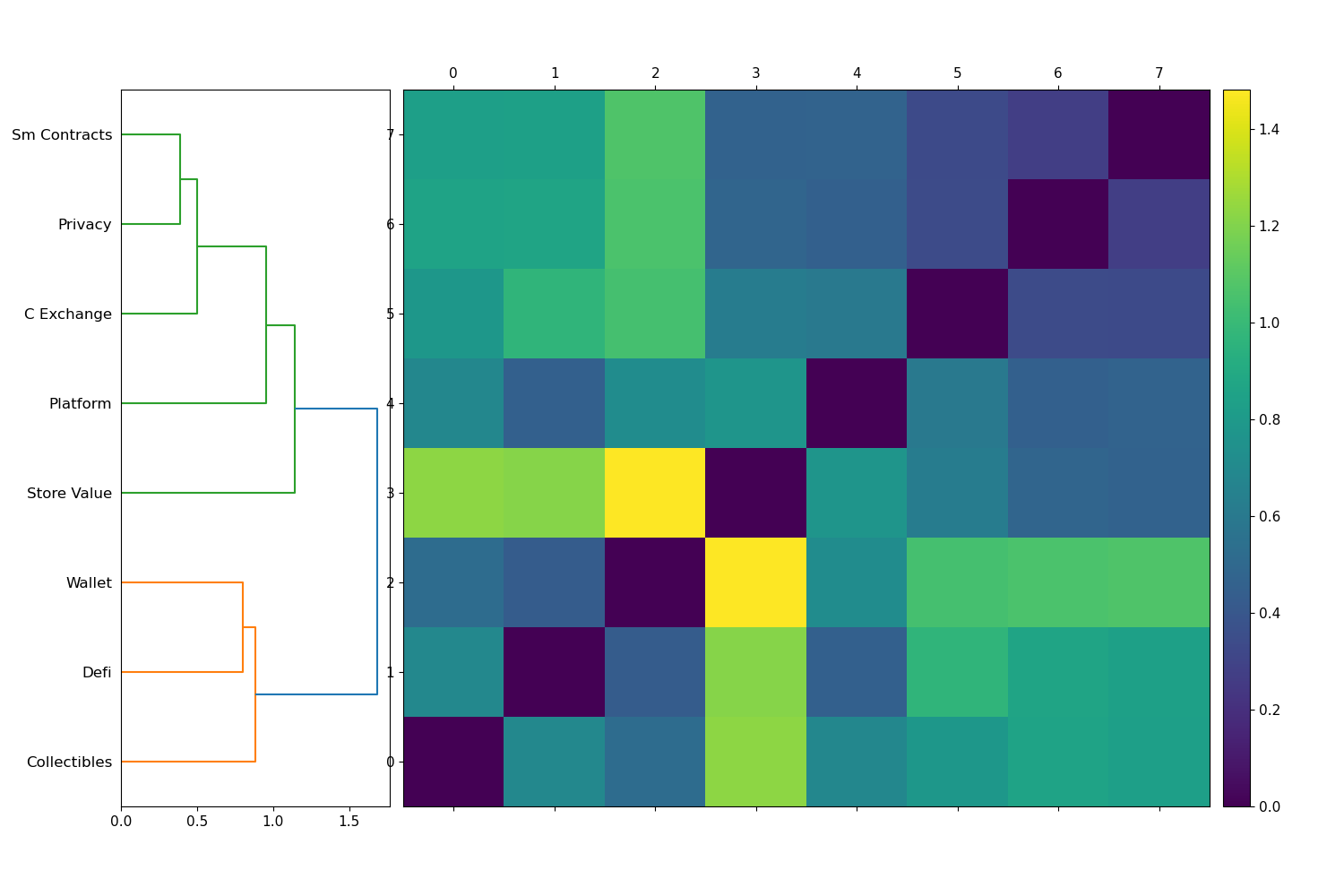}
        \caption{}
        \label{fig:Adaptspec_crypto_dendrogram}
    \end{subfigure}
    \begin{subfigure}[b]{0.75\textwidth}
        \includegraphics[width=\textwidth]{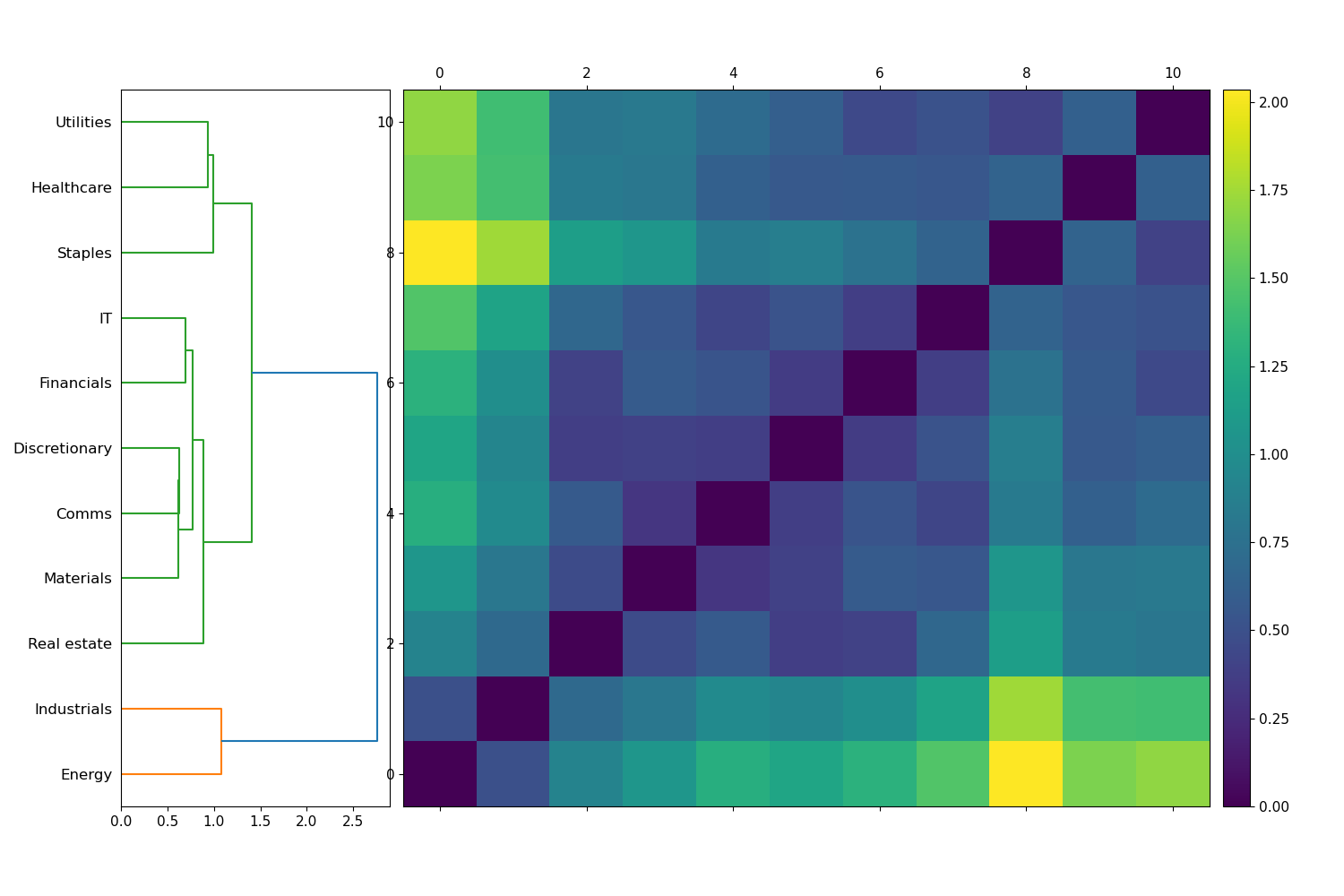}
        \caption{}
        \label{fig:Adaptspec_equity_dendrogram}
    \end{subfigure}
    \caption{Hierarchical clustering applied to cryptocurrency and equity time-varying spectrum distance matrices, $D^{c, (\nu, t_1)}$ and $D^{e, (\nu, t_2)}$.}
    \label{fig:Dendrogram_adaptspec_surface}
\end{figure}

\begin{figure}
    \centering
    \begin{subfigure}[b]{0.48\textwidth}
        \includegraphics[width=\textwidth]{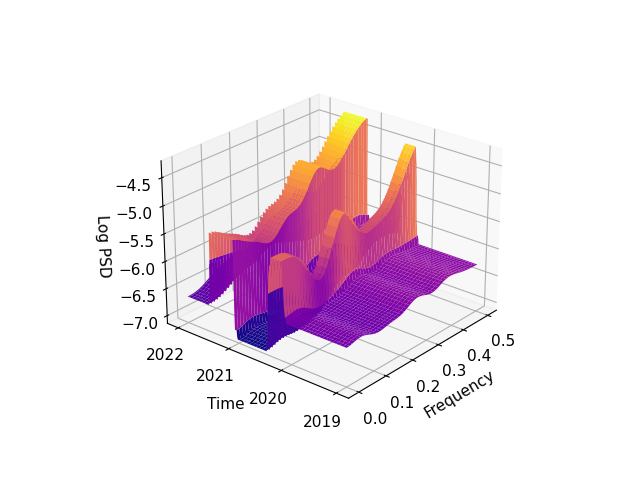}
        \caption{}
        \label{fig:TVS_privacy}
    \end{subfigure}
    \begin{subfigure}[b]{0.48\textwidth}
        \includegraphics[width=\textwidth]{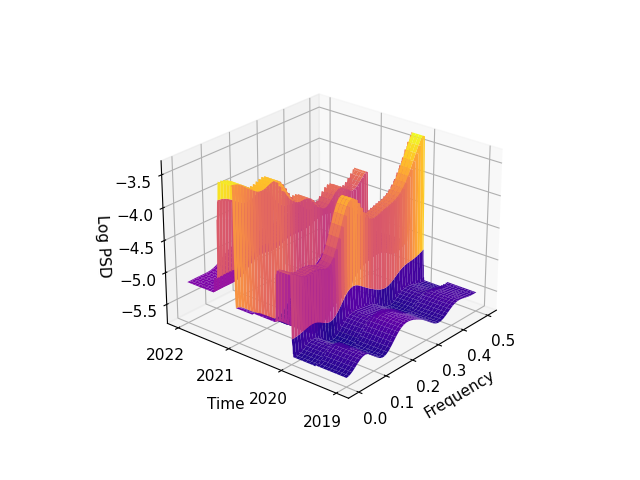}
        \caption{}
        \label{fig:TVS_Wallet}
    \end{subfigure}
    \begin{subfigure}[b]{0.48\textwidth}
        \includegraphics[width=\textwidth]{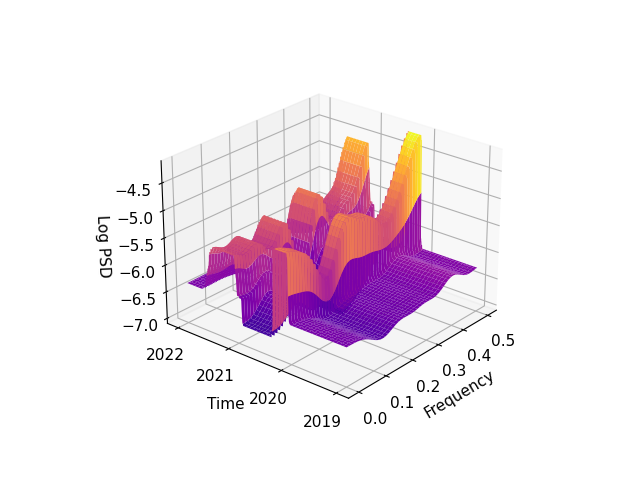}
        \caption{}
        \label{fig:TVS_smart_contracts}
    \end{subfigure}
    \begin{subfigure}[b]{0.48\textwidth}
        \includegraphics[width=\textwidth]{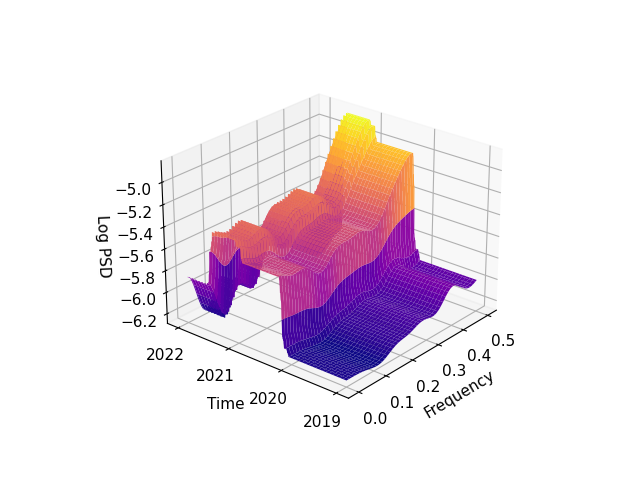}
        \caption{}
        \label{fig:TVS_platform}
    \end{subfigure}
    \caption{A sample of cryptocurrency evolutionary power spectra (a) Privacy (b) Wallet, (c) Smart Contracts, (d) Platform.}
    \label{fig:Crypto_TVS}
\end{figure}

\begin{figure}
    \centering
    \begin{subfigure}[b]{0.48\textwidth}
        \includegraphics[width=\textwidth]{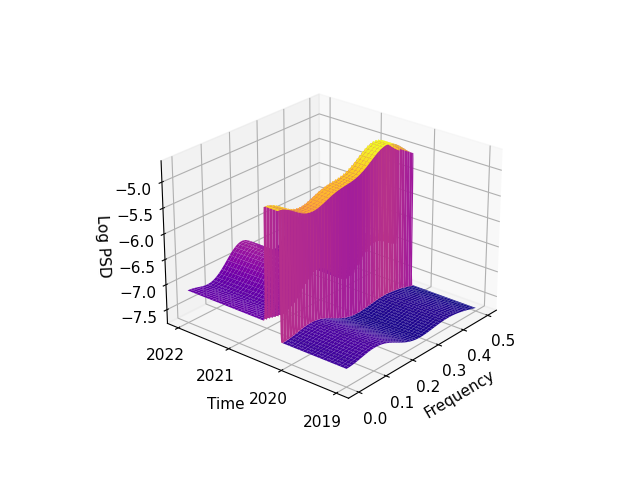}
        \caption{}
        \label{fig:TVS_energy}
    \end{subfigure}
    \begin{subfigure}[b]{0.48\textwidth}
        \includegraphics[width=\textwidth]{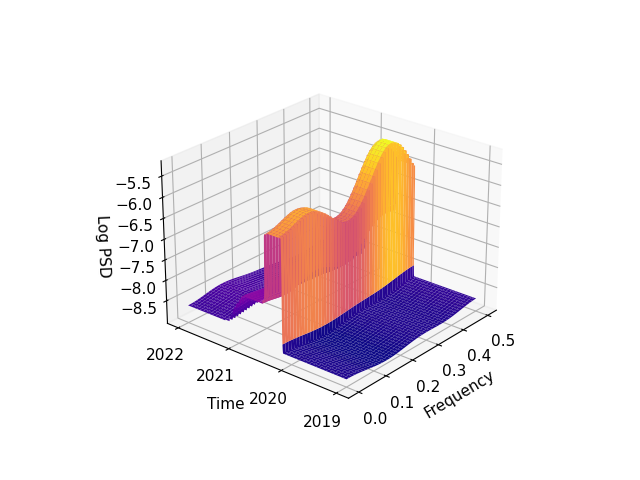}
        \caption{}
        \label{fig:TVS_financials}
    \end{subfigure}
    \begin{subfigure}[b]{0.48\textwidth}
        \includegraphics[width=\textwidth]{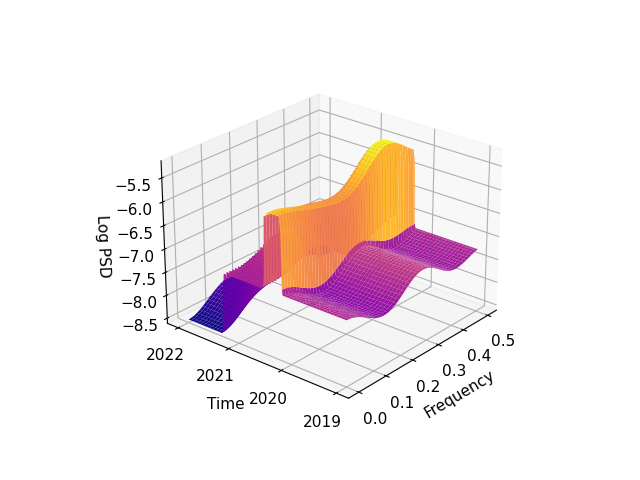}
        \caption{}
        \label{fig:TVS_industrials}
    \end{subfigure}
    \begin{subfigure}[b]{0.48\textwidth}
        \includegraphics[width=\textwidth]{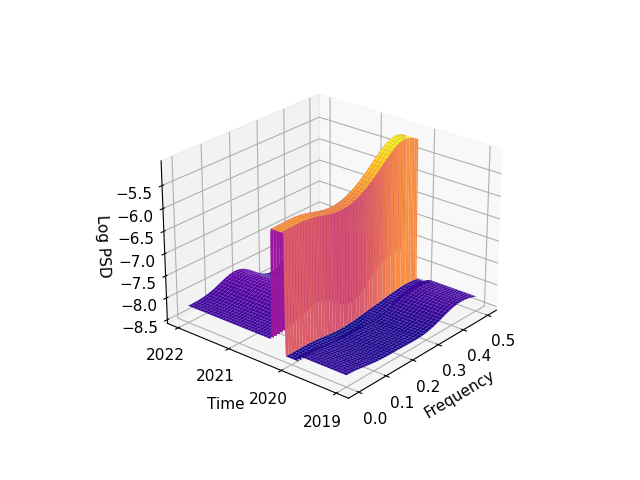}
        \caption{}
        \label{fig:TVS_materials}
    \end{subfigure}
    \caption{A sample of equity evolutionary power spectra (a) Energy (b) Financials, (c) Industrials, (d) Materials.}
    \label{fig:Equity_TVS}
\end{figure}

The time-varying power spectral densities for the cryptocurrency and equity collections are shown in Figures \ref{fig:Crypto_TVS} and \ref{fig:Equity_TVS}. Their surfaces reveal some interesting differences in the evolutionary behaviour of cryptocurrency and equity sectors. Among the four cryptocurrency sectors shown, all exhibit highly varying power spectra over time. In general, there is an overwhelming tendency for higher frequency components to dominate, indicating the abrupt changes in behaviours exhibited by cryptocurrencies. This is particularly pronounced in early 2020, with all sectors showing a significant translation upward in the level of the power spectrum, and a clear shift toward higher frequency components exhibiting most power. This is indicative of the volatility experienced in the market during the COVID-19 market crash. These trends are relatively consistent in Figures \ref{fig:TVS_privacy}, \ref{fig:TVS_Wallet}, \ref{fig:TVS_smart_contracts} and \ref{fig:TVS_platform} representing privacy, wallet, smart contract and platform focused cryptocurrencies. 

Next we turn to our collection of equities. Equity sectors are shown in Figure \ref{fig:Equity_TVS}, with energy, financials, industrials and materials' time-varying power spectra shown in Figures \ref{fig:TVS_energy}, \ref{fig:TVS_financials}, \ref{fig:TVS_industrials} and \ref{fig:TVS_materials} respectively. The power spectra exhibited by equities is significantly smoother, and there are clearly less significant re-allocations of power among frequency components when transitioning to adjacent segments. Like the cryptocurrency surfaces, the equity spectra exhibit a pronounced translation during the COVID-19 market crash, with significant power moving to higher frequency components among all equity sectors. 

The clustered distance matrix between the time-varying spectra of respective sectors for the cryptocurrency and equity collections are shown in Figure \ref{fig:Dendrogram_adaptspec_surface}. Turning first to cryptocurrencies, shown in Figure \ref{fig:Adaptspec_crypto_dendrogram}, two characteristic clusters of sector behaviours emerge. Cluster one consists of smart contracts, privacy centralized exchange, platform and store of value themed cryptocurrency sectors. The second cluster consists of wallet, decentralized finance and collectibles/NFT-themed cryptocurrency sectors. The overall cluster structure in evolutionary spectra among cryptocurrency sectors is shown to be relatively diffuse, without the emergence of a predominant collection of sectors with characteristic spectral properties. By contrast, the similarity in time-varying power spectra of equity sectors shown in Figure \ref{fig:Adaptspec_equity_dendrogram}, highlights the emergence of a dominant cluster, consisting of most equity sectors. The primary cluster consists of: Utilities, Healthcare, Consumer staples, Information technology, Financials, Consumer discretionary, Communications, Materials and Real estate. All sectors exhibit power spectra with similar levels of amplitude, and clear peaks in the power spectra at high frequency components. In the outlier cluster, which consists of the Industrials and Energy sectors, spectral peaks are exhibited at slightly lower frequency components - differentiating the surfaces from the rest of the collection. Viewing both dendrograms in conjunction with the time-varying power spectra figures, more consistent similarity is seen among equity sector behaviours. The time-varying power spectra of cryptocurrencies and the increased frequency of structural breaks reflects the higher levels of volatility exhibited by the underlying cryptocurrency sectors.

 
 
 

\section{Cryptocurrency time-varying optimization}
\label{Cryptocurrency_time_varying_optimization}

In this section we introduce two new simulation frameworks to test the relative importance of security selection and sector allocation in the context of cryptocurrency portfolio management. There is a litany of research in the quantitative finance and optimization fields focused on untangling this question for more traditional asset classes. However, given the nascence of the cryptocurrency market, it is worth investigating the most purposeful decisions for candidate portfolio managers. 

Specifically, we wish to determine whether there is more portfolio alpha available at the stock selection, or sector allocation level. To test our hypothesis, we implement two iterative algorithms which aim to maximize portfolio Sharpe ratio subject to various portfolio constraints. 

\begin{algorithm}[H]
	\caption{Optimal security selection} 
	\label{algorithm}
	\begin{algorithmic}[1]	
	    \State Choose a smoothing rate $S$ over time.
	    \State Choose a number of `best' securities to sample from each sector, $B$.
	    \State Let the cryptocurrency sectors be indexed $\psi_l: l=1,...,L$.
	    \State Let the number of securities per cryptocurrency sector be $J=7$.
	    \For{$t_1=S \,\text{ to }\, T_1$}
	    \For{$\psi_l: l=1 \,\text{ to }\, L $}
	    \For{$j=1 \,\text{ to }\, J$}
	    \State Compute total log returns $Z^{\psi_l}_j(t_1) = \sum^{t_1}_{d=t_1-S} R^{\psi_l}_j(d)$.
	    \State Compute volatility $\sigma_j^{2,\psi_l}(t_1)$.
	    \State Compute risk-adjusted return $RAR^{\psi_l}_j (t_1) = {Z^{\psi_l}_j(t_1)} / {\sigma_j^{2,\psi_l}(t_1)}$
	    \EndFor 
	    \State Select $B_{\psi_l}$ best securities within sector $\psi_l$ ranked by $RAR_j^{\psi_l}(t_1)$.
	    \State Compute sector return $Z^{\psi_l}(t_1) = \frac{1}{B_{\psi_l}} \sum^{B_{\psi_l}}_{j=1} Z^{\psi_l}_j (t_1)$
	    \State Append sector return $Z^{\psi_l}(t_1)$ .
	    \EndFor 
	    \State Compute portfolio return $Z(t_1) = \frac{1}{L} \sum^L_{l=1} Z^{\psi_l} (t_1)$.
	    \State Append total portfolio return $Z(t_1)$.
	    \EndFor 
	    \State Compute optimal cryptocurrency portfolio return over all time $R^{o,c}_p = \sum^{T_1}_{t_1=S} Z(t_1)$.
	\end{algorithmic}
\end{algorithm}

In Algorithm 1, we implement a time-varying simulation where we select a collection of cryptocurrency securities from each sector that exhibit a maximal Sharpe ratio over some trailing window of size $S$. In doing so, we allocate uniform weight across our 8 candidate cryptocurrency sectors, and choose a sample of the $B$ best securities within each sector, measured by their trailing volatility-adjusted log returns. To simulate the dynamics of a portfolio management operation, there is no optimization routine performed at the sector-specific level, and all sectors are assumed to have the same amount of allocated capital at all points in time $t_1$. This is consistent with the concept of daily portfolio rebalancing.

\begin{algorithm}[H]
	\caption{Optimal sector allocation} 
	\label{algorithm}
	\begin{algorithmic}[1]	
	    \State Choose a smoothing rate $S$ over time.
	    \State Choose a `best' number of sectors, $B$, to allocate capital to.
	    \State Let the cryptocurrency sectors be indexed $\psi_l: l=1,...,L$.
	    \State Let the number of securities per cryptocurrency sector be $J=7$.
	    \For{$t_1=S \,\text{ to }\, T_1$}
	    \For{$\psi_l: l=1 \,\text{ to }\, L $}
	    \For{$j=1 \,\text{ to }\, J$}
	    \State Compute total log returns for each security $Z_j^{\psi_l} (t_1) = \sum^{t_1}_{d=t_1-S} R^{\psi_l}_j(d)$
	    \EndFor
	    \State Assume uniform weight for each security $w_j^{\psi_l} = 1/J \quad \forall j \in \{1,...,J\}$
	    \State Compute sector returns $Z^{\psi_l}(t_1) = \sum^{J}_{j=1} Z^{\psi_l}_j (t_1) w_j^{\psi_l}$
	    \State Compute sector volatility $\sigma^{2, \psi_l} (t_1) = \mathbf{w}^{\psi_l T} \Sigma(t_1) \mathbf{w}^{\psi_l}$
	    \State Compute sector risk-adjusted return $RAR^{\psi_l} (t_1) = Z^{\psi_l} (t_1) / \sigma^{2, \psi_l} (t_1)$
	    \EndFor
	    \State Select $B$ best sectors ranked by $RAR^{\psi_l}(t_1)$
	    \State Compute total portfolio return $Z(t_1) = \frac{1}{B} \sum^{B}_{q=1} Z^{\psi_l}_q (t_1)$
	    \State Append total portfolio return $Z(t_1)$
	    \EndFor
	    \State Compute portfolio return over time $R^{o,s}_p = \sum^{T_1}_{t_1=S} Z(t_1)$.
	\end{algorithmic}
\end{algorithm}

In Algorithm 2, we seek to determine the $B$ best performing cryptocurrency sectors over some trailing window $S$, and uniformly allocate capital across these sectors. To judge the rolling performance of each sector, we compute the sector risk-adjusted returns at each point in time $t_1$ (assuming uniform weight among constituent securities), and again select the $B$ best sectors to allocate capital toward. In this simulation, we have a constant assumption of uniform weight allocated across cryptocurrency securities within any candidate sector, and uniform weight allocated across the $B$ best sectors. This is done to explicitly test for alpha generation at the level of sector allocation. 

\begin{table}
\centering
\begin{tabular}{ |p{1.5cm}|p{1.5cm}|p{1.5cm}|p{1.5cm}|}
 \hline
 \multicolumn{4}{|c|}{Cryptocurrency portfolio return simulations } \\
 \hline
 $S$ & $B$ & $R^{o,c}_p$ & $R^{o,s}_p$  \\
 \hline
 120 & 2 & 215\% & 177\% \\
 120 & 3 & 210\% & 201\% \\
 120 & 4 & 210\% & 219\% \\
 120 & 5 & 214\% & 235\% \\
  150 & 2 & 155\% & 130\% \\
 150 & 3 & 167\% & 162\% \\
 150 & 4 & 177\% & 207\% \\
 150 & 5 & 177\% & 208\% \\
 180 & 2 & 166\% & 168\% \\
 180 & 3 & 149\% & 179\% \\
 180 & 4 & 179\% & 175\% \\
 180 & 5 & 187\% & 180\% \\
\hline
\end{tabular}
\caption{Simulated cryptocurrency portfolio return simulations. In various iterations of the simulation, $B$ and $S$ are varied. Our results are inconclusive as to whether there is more value in cryptocurrency sector allocation or security selection.}
\label{tab:Optimal_cryptocurrency_portfolios}
\end{table}

We run Algorithm 1 and Algorithm 2 for a range of values $S$ and $B$, and compute the portfolio returns for optimal cryptocurrency selection $R_p^{o,c}$ and optimal security selection $R_p^{o,s}$ for all possible parameter combinations. The results are shown in Table \ref{tab:Optimal_cryptocurrency_portfolios}, and are relatively stable and consistent among all possible parameter values. The table highlights several interesting findings. The most notable observation however, is that there is no clear winner when comparing the realised performance of the security selection (Algorithm 1) and sector allocation (Algorithm 2) algorithms. The second notable observation is the significantly improved performance when $S=120$. In both Algorithm 1 and Algorithm 2, the highest average portfolio returns are exhibited with a smaller rolling window. This reflects the dynamic nature of cryptocurrency behaviours, and demonstrates the need for continued monitoring of market dynamics such as correlation. The final interesting observation, is that there is typically an improvement in portfolio performance (subject to a constant level $S$) as $B$ increases. This phenomenon occurs in both portfolios, but is particularly pronounced in the sector  allocation simulation. There are two takeaways from this observation. First, there is clearly a meaningful benefit when cryptocurrency portfolios are well diversified. The second point is more subtle. Given that the improvement (subject to a constant $S$) when $B$ increases is more pronounced in Algorithm 2, it highlights a potential risk in cryptocurrency characteristic behaviours. There appears to be a tendency for certain cryptocurrency sectors to trend, and holding more exposure to a variety of cryptocurrency sectors could be of great benefit to any investor in the cryptocurrency markets. This finding may question the notion that holding a major (dominant) cryptocurrency such as Bitcoin or Ethereum, provides satisfactory exposure to the entire cryptocurrency market.

\section{Conclusion}
\label{Conclusion}

In Section \ref{RMT} we use random matrix theory in a time-varying capacity to explore the evolutionary properties of cryptocurrency and equity correlation matrices. Our methodology produces several interesting findings. First, the equity collection consistently exhibits a larger number of non-random eigenvalues over our analysis window. This is consistent with the findings shown in \cite{Drod2020}. By contrast with the equity market, we show that the cryptocurrency market consistently displays a smaller number of non-random eigenvalues. When interpreting this distinction between the two asset classes, it is worth clarifying the possible interpretation of such findings. Large outliers in the eigenspectrum between the bulk of the distribution of eigenvalues and the largest eigenvalue (representing the strength of collective behaviour, or the market) measures the degree of collectivity in groups of sectors. The origins of such collectivity in the sectors could stem from exogenous or endogenous factors. In the case of the former, this may represent new laws or regulations which only impact one industry for example, and in the case of the latter, investor market expectations may shift with respect to only one sector for some sentiment-driven investor reasoning. Our analysis certainly implies that there is a greater degree of complexity in the eigenspectrum exhibited by equities, and this is consistent over time. The second key finding is the notable inverse relationship observed between the magnitude of the first correlation matrix eigenvalue and the number of non-random eigenvalues based on the random matrix theory. This relationship implies that when the collective strength of market behaviours increases (associated with the magnitude of the first eigenvalue), the effect of the market itself becomes more pronounced on underlying financial market dynamics. 

In Section \ref{Evolutionary_dynamics_sector}, we demonstrate marked differences in the similarity of cryptocurrency and equity sectors' evolutionary dynamics. In particular, cryptocurrencies exhibit quite an obvious `coupling' behaviour, where sectors with similar underlying dynamics are revealed. By contrast, hierarchical clustering on our equity collection reveals one predominant cluster of similar behaviours, and a small group of outlier sectors (Utilities, Energy and Financials). As the market continues to develop, such techniques could be used to identify sub-groups within the cryptocurrency space that behave most similarly. Given the nascence and rapid changes in the market, this is worth monitoring closely.

In Section \ref{Structural_breaks_non_stationarity} we study the collective similarity in structural break propagation and time-varying power spectra in both collections. Our frameworks reveal significantly greater similarity in the structural break dynamics of equity sectors than that of cryptocurrency sectors. This is consistent with prior research which demonstrates the existence of erratic behaviours in cryptocurrency returns and volatility \cite{James2021_crypto}. Our experiments also reveal that cryptocurrency sectors tend to consist of a greater number locally-stationary segments than equities, further highlighting the tendency for changes in time series behaviour (structural breaks) in cryptocurrency sectors.  To better understand the subtle difference in nonstationarity and long range dependence, we study the similarity between sectors' time-varying power spectra. Our experiments reveal that equities produce one predominant cluster of evolutionary spectra with just two outliers (Industrials and Energy), while cryptocurrencies are partitioned into two, (approximately) equally-sized clusters.

Finally in Section \ref{Cryptocurrency_time_varying_optimization} we focus exclusively on the burgeoning cryptocurrency market, and investigate the relative importance of security selection and sector allocation in cryptocurrency portfolio management. We implement two algorithms, with a range of values for our smoothing window parameter $S$, and best number of samples $B$ (where we sample underlying securities in Algorithm 1 and cryptocurrency sectors in Algorithm 2). Our experiments reveal that both decisions are crucial in cryptocurrency portfolio management, and neither piece is of greater importance than the other. Optimal portfolio returns are realised with the smallest possible smoothing window ($S=120$), which demonstrates the dynamic nature of cryptocurrency behaviours and the need for adaptability in portfolio management decision-making. Finally, we demonstrate (in both algorithms), the benefit of diversifying across securities and sectors. Our experiments reveal that subject to a candidate level of $S$, as $B$ increases portfolio returns also tend to increase.

We hope that this paper may serve as a current accompaniment to academics and practitioners seeking to understand the evolution of individual and collective cryptocurrency behaviours, and the implications for managing a portfolio of cryptocurrency investments. However, there are several avenues to further develop upon the frameworks introduced in this paper. First, one could extend all the analysis in this paper to other asset classes such as bonds, currencies and commodities. The study of time-varying sector explanatory variance could be accompanied by a study of sector equity returns. One could test the `resilience' or `elasticity' of sector returns to spikes in collective movements (typically associated with market crises). The study of sector structural breaks and evolutionary power spectra could be extended. In this work, we choose one representative security as a proxy for each sector. A focused paper on this particular issue could study intra-sector similarity in addition to inter-sector similarity. This may reveal specific sectors that exhibit more self-similarity in their structural break profile. One could also test for the consistency in collective similarity between evolutionary spectra and structural breaks, among various asset classes. Such analysis could identify anomalous sectors with regards to their evolutionary power spectra, or structural break profile. Finally, one could further develop the trading strategy simulations by incorporating other portfolio management decisions, or treating the problem as a more explicit, higher dimensional matter - where one could learn model parameters such as $S$ and $B$ subject to some candidate loss function. 


\section{Acknowledgments}
\noindent The author would like to thank Kevin Chin from Arowana \& Co. who helped provide some of the key motivating questions for the author. 

\appendix

\section{Mathematical objects}
\label{appendix:MathematicalNotation}

\begin{center}
\begin{longtable}{|l|l|}
\caption{Mathematical objects} \label{tab:Mathematical_objects} \\
\hline \multicolumn{1}{|c|}{\textbf{Mathematical object}} & \multicolumn{1}{c|}{\textbf{Description}} \\ \hline 
\endfirsthead
\multicolumn{2}{c}%
{{\bfseries \tablename\ \thetable{} -- continued from previous page}} \\
\hline \multicolumn{1}{|c|}{\textbf{Mathematical object}} & \multicolumn{1}{c|}{\textbf{Description}} \\ \hline 
\endhead

\hline \multicolumn{2}{|r|}{{Continued on next page}} \\ \hline
\endfoot

\hline \hline
\endlastfoot

  $t_1 = 1,...,T_1$ & Cryptocurrency time index, $T_1 = 1065$\\
  $t_2 = 1,...,T_2$ & Equity time index, $T_2 = 733$\\
  $p^c_i$ & Multivariate time series of cryptocurrency prices. \\
  $p^e_j$ & Multivariate time series of equity prices. \\
  $R^c_i$ & Multivariate time series of cryptocurrency log returns. \\
  $R^e_j$ & Multivariate time series of equity log returns. \\
  $N$ & Number of cryptocurrency time series. \\
  $K$ & Number of equity time series. \\
  $\Omega^c$ & Cryptocurrency correlation matrix. \\
  $\Omega^e$ & Equity correlation matrix. \\
$\lambda_1^c,...,\lambda^c_N$ & Cryptocurrency eigenvalues ordered by magnitude. \\
  $\lambda_1^e,...,\lambda^e_K$ & Equity eigenvalues ordered by magnitude. \\
  $\tilde{\lambda}_1^c,...,\tilde{\lambda}^c_N$ & Normalized cryptocurrency eigenvalues ordered by magnitude. \\
  $\tilde{\lambda}_1^e,...,\tilde{\lambda}^e_K$ & Normalized equity eigenvalues ordered by magnitude. \\
    $p(\lambda^c)$ & Probability density function of cryptocurrency eigenvalues. \\
  $p(\lambda^e)$ & Probability density function of equity eigenvalues. \\
  $l$ & Cryptocurrency sector time series index, $L = 8$. \\
  $p$ & Equity sector time series index, $P=11$. \\
  $\Omega^{c,\psi_l}$ & Cryptocurrency correlation matrix for sector $l$. \\
  $\Omega^{e,\psi_p}$ & Equity correlation matrix for sector $p$. \\
  $\tilde{\lambda_1}^{c,\psi_l}$ & Cryptocurrency dominant eigenvalue explanatory variance for sector $l$. \\
  $\tilde{\lambda_1}^{e,\psi_p}$ & Equity dominant eigenvalue explanatory variance for sector $p$. \\
  $D^{c,\tilde{\lambda_1}}$ & Cryptocurrency dominant eigenvalue distance matrix (collective behaviour). \\
  $D^{e,\tilde{\lambda_1}}$ & Equity dominant eigenvalue distance matrix (collective behaviour). \\
  $D^{c, MJW}$ & Cryptocurrency MJ-Wasserstein distance matrix (structural breaks). \\
  $D^{e, MJW}$ & Equity MJ-Wasserstein distance matrix (structural breaks). \\
   $f_{\psi_l}^c (\nu, t_1)$ & Cryptocurrency time-varying power spectrum for sector $l$. \\
   $f_{\psi_p}^e (\nu, t_2)$ & Equity time-varying power spectrum for sector $p$. \\
   $D^{c,(\nu, t_1)}$ & Cryptocurrency time-varying power spectrum distance matrix. \\
   $D^{e,(\nu, t_2)}$ & Equity time-varying power spectrum distance matrix. \\

\end{longtable}
\end{center}

\section{Securities}
\label{appendix:Securities}

\begin{center}
\begin{longtable}{|l|l|l|}
\caption{Cryptocurrency securities.} \label{tab:Cryptocurrency} \\
\hline \multicolumn{1}{|c|}{\textbf{Security}} & \multicolumn{1}{c|}{\textbf{Ticker}} & \multicolumn{1}{c|}{\textbf{Sector}} \\ \hline 
\endfirsthead
\multicolumn{3}{c}%
{{\bfseries \tablename\ \thetable{} -- continued from previous page}} \\
\hline \multicolumn{1}{|c|}{\textbf{Security}} & \multicolumn{1}{c|}{\textbf{Ticker}} & \multicolumn{1}{c|}{\textbf{Sector}} \\ \hline 
\endhead

\hline \multicolumn{3}{|r|}{{Continued on next page}} \\ \hline
\endfoot

\hline \hline
\endlastfoot

  Binance Coin & BNB & Centralized exchange \\
  Bankera & BNK & Centralized exchange \\
 Crypto.com & CRO & Centralized exchange \\
 Huobi Token & HT & Centralized exchange \\
 KuCoin Token & KCS & Centralized exchange \\
 LATOKEN & LA & Centralized exchange \\
ZB Token & ZB & Centralized exchange \\
Enjin Coin & ENJ & Collectibles/NFTs \\
Decentraland & MANA & Collectibles/NFTs \\
Phantasma & SOUL & Collectibles/NFTs \\
 Syscoin & SYS & Collectibles/NFTs \\
 THETA & THETA & Collectibles/NFTs \\
 WAXP & WAXP & Collectibles/NFTs \\
 Tezos & XTZ & Collectibles/NFTs \\
 Basic Attention Token & BAT & Decentralized finance \\
 Bancor & BNT & Decentralized finance \\
 Fantom & FTM & Decentralized finance \\
 Chainlink & LINK & Decentralized finance \\
 Loopring & LRC & Decentralized finance \\
 Maker & MKR & Decentralized finance \\
 Tezos & XTZ & Decentralized finance \\
 Cardano & ADA & Platform \\
 Ethereum Classic & ETC & Platform \\
 Fantom & FTM & Platform \\
 Chinlink & LINK & Platform \\
 Decentraland & MANA & Platform \\
 Neo & NEO & Platform \\
Tezos & XTZ & Platform \\
 Decred & DCR & Privacy \\
 Flux & FLUX & Privacy \\
 iExec RLC & RLC & Privacy \\
 Monero & XMR & Privacy \\
 Verge & XVG & Privacy \\
 Zcash & ZEC & Privacy \\
 Horizen & ZEN & Privacy \\
 Cardano & ADA & Smart Contracts \\
 Binance Coin & BNB & Smart Contracts \\
 Ethereum Classic & ETC & Smart Contracts \\
 Ethereum & ETH & Smart Contracts \\
 Chainlink & LINK & Smart Contracts \\
 VeChain & VET & Smart Contracts \\
 Stellar & XLM & Smart Contracts \\
 Bitcoin Cash & BCH & Store of Value \\
 Bitcoin SV & BSV & Store of Value \\
 Bitcoin & BTC & Store of Value \\
 Decred & DCR & Store of Value \\
 Maker & MKR & Store of Value \\
 Bitcoin Plus & XBC & Store of Value \\
 Nano & XNO & Store of Value \\
 Bread & BRD & Wallet \\
 Divi & DIVI & Wallet \\
 Electroneum & ETN & Wallet \\
 STASIS EURO & EURS & Wallet \\
 Metronome & MET & Wallet \\
 Pillar & PLR & Wallet \\
 Voyager Token & VGX & Wallet \\
\end{longtable}
\end{center}

\begin{center}
\begin{longtable}{|l|l|l|}
\caption{Equity securities.} \label{tab:Equities} \\
\hline \multicolumn{1}{|c|}{\textbf{Security}} & \multicolumn{1}{c|}{\textbf{Ticker}} & \multicolumn{1}{c|}{\textbf{Sector}} \\ \hline 
\endfirsthead
\multicolumn{3}{c}%
{{\bfseries \tablename\ \thetable{} -- continued from previous page}} \\
\hline \multicolumn{1}{|c|}{\textbf{Security}} & \multicolumn{1}{c|}{\textbf{Ticker}} & \multicolumn{1}{c|}{\textbf{Sector}} \\ \hline 
\endhead

\hline \multicolumn{3}{|r|}{{Continued on next page}} \\ \hline
\endfoot

\hline \hline
\endlastfoot

  The Walt Disney Co. & DIS & Communication Services \\
  Alphabet Inc. & GOOG & Communication Services \\
  The Interpublic Group & IPG & Communication Services \\
  Lumen Technologies & LUMN & Communication Services \\
  Netflix Inc. & NFLX & Communication Services \\
  News Corp. & NWSA & Communication Services \\
  Verizon Communications & VZ & Communication Services \\
  Amazon.com & AMZN & Consumer Discretionary \\
  AutoZone Inc. & AZO & Consumer Discretionary \\
  Best Buy Co. & BBY & Consumer Discretionary \\
  General Motors & GM & Consumer Discretionary \\
  Home Depot & HD & Consumer Discretionary \\
  NIKE Inc. & NKE & Consumer Discretionary \\
  Starbucks Corp. & SBUX & Consumer Discretionary \\
  Kellogg Co. & K & Consumer Staples \\
  The Kraft Heinz Co. & KHC & Consumer Staples \\
  The Coca-Cola Co. & KO & Consumer Staples \\
  PepsiCo. Inc & PEP & Consumer Staples \\
  Procter \& Gamble Co. & PG & Consumer Staples \\
  Tyson Foods Inc. & TSN & Consumer Staples \\
  Walmart Inc. & WMT & Consumer Staples \\
  Chevron cO. & CVX & Energy \\
  Haliburton Co. & HAL & Energy \\
  Hess Corp. & HES & Energy \\
  Kinder Morgan Inc. & KMI & Energy \\
  Schlumberger NV & SLB & Energy \\
  Valero Energy Corp. & VLO & Energy \\
  Exxon Mobil Corp. & XOM & Energy \\
  American Express Co. & AXP & Financials \\
  Bank of America Corp. & BAC & Financials \\
  Citigroup & C & Financials \\
  JPMorgan Chase \& Co. & JPM & Financials \\
  MSCI Inc. & MSCI & Financials \\
  Wells Fargo \& Co. & WFC & Financials \\
  Willis Towers Watson PLC & WLTW & Financials \\
  Amgen Inc. & AMGN & Healthcare \\
  Baxter International Inc. & BAX & Healthcare \\
  Danaher Corp. & DHR & Healthcare \\
  Johnson \& Johnson & JNJ & Healthcare \\
  McKesson Corp. & MCK & Healthcare \\
  Merck \& Co. Inc. & MRK & Healthcare \\
  Pfizer Inc. & PFE & Healthcare \\
  The Boeing Co. & BA & Industrials \\
  Caterpillar Inc. & CAT & Industrials \\
  Deere \& Co. & DE & Industrials \\
  Equifax Inc. & EFX & Industrials \\
  FedEx Corp. & FDX & Industrials \\
  General Electric Co. & GE & Industrials \\
  3M Co. & MMM & Industrials \\
  Autodesk Inc. & ADSK & Information Technology \\
  Broadcom Inc. & AVGO & Information Technology \\
  Cisco Systems Inc. & CSCO & Information Technology \\
  Citrix Systems & CTXS & Information Technology \\
  Hewlett Packard Enterprise Co. & HPE & Information Technology \\
  Microsoft Corp. & MSFT & Information Technology \\
  NortonLifeLock Inc. & NLOK & Information Technology \\
  Ball Corp & BLL & Materials \\
  DuPont de Nemours Inc. & DD & Materials \\
  FMC Corp. & FMC & Materials \\
  International Paper Co. & IP & Materials \\
  LyondellBasell Industries NV & LYB & Materials \\
  Vulcan Materials Co. & VMC & Materials \\
  Westrock Co. & WRK & Materials \\
  American Tower Corp. & AMT & Real estate \\
  Boston Properties Inc. & BXP & Real estate \\
  CBRE Group Inc. & CBRE & Real estate \\
  Essex Property Trust Inc. & ESS & Real estate \\
  Extra Space Storage Inc. & EXR & Real estate \\
  Public Storage & PSA & Real estate \\
  Welltower Inc. & WELL & Real estate \\
  American Electric Power Co. Inc. & AEP & Utilities \\
  American Water Works Co. & AWK & Utilities \\
  Duke Energy Corp. & DUK & Utilities \\
  Exelon Corp. & EXC & Utilities \\
  Alliant Energy Corp. & LNT & Utilities \\
  Pinnacle West Capital Corp. & PNW & Utilities \\
  WEC Energy Group Inc. & WEC & Utilities \\
\end{longtable}
\end{center}

\section{Bayesian change point detection algorithm}
\label{appendix:RJMCMCsampling}

In this section, we describe the Bayesian change point detection algorithm that was used in Section \ref{Structural_breaks_non_stationarity} of this paper. We  use a reversible jump Markov chain Monte Carlo (RJMCMC) algorithm \cite{Rosen2017} that continues to partition a (presumably nonstationary) time series into $m$ segments, where $m$ also changes with time. The sampling scheme produces a distribution over possible models $m$ and also produces a set of $m$ change points.

The method identifies structural breaks/change points based on changes in the evolutionary power spectral density \cite{Rosen2012}. Many change point detection methodologies that exist in the time domain fail to identify suitable change points when time series exhibit dependence. This is commonly the case with autoregressive processes, which often display similar patterns to that exhibited by financial time series. Previous research \cite{Rosen2017} has highlighted that the methodology used in this paper can detect changes in a sequence of autoregressive processes, highlighting the algorithm's ability to partition data that exhibits dependence.

We follow Rosen et al. and James and Menzies \cite{Rosen2012, james2021_spectral} in our implementation of the RJMCMC sampling scheme. In particular, James and Menzies introduce a framework where an optimal number of basis functions is used when estimating a candidate power spectral density. We denote a varying partition of the time series by $\bs{\xi}_{m} = (\xi_{0,m},...,\xi_{m,m})$; these are our $m$ change points (excluding $\xi_{m,m}$, which by convention is always going to be the final time point). The algorithm must consider a vector of \emph{amplitude parameters} $\bs{\tau}_{m}^{2} = (\tau_{1,m}^{2},...,\tau_{m,m}^{2})'$ and \emph{regression coefficients} $\bs{\beta}_{m} = (\bs{\beta'}_{1,m},...,\bs{\beta'}_{m,m})$ that we wish to estimate, for the $j$th component within a partition of $m$ segments, $j=1,...,m.$ For notational convenience, $\bs{\beta}_{j,m}, j=1,...,m,$ is assumed to include the first entry, $\alpha_{0j,m}.$ In the sections that follows, the superscripts $c$ and $p$ refer to the current and proposed values in the sampling scheme, respectively. This is consistent with most notation used in Markov chain Monte Carlo algorithms.

We begin by describing the \textbf{between-model moves}, where the number of change points $m$ may be changed. Let $\bs{\theta}_{m} = (\bs{\xi}'_{m}, \bs{\tau}^{2'}_{m}, \bs{\beta'}_{m})$ be the model parameters at some point in the RJMCMC sampling scheme and assume that the chain starts at $(m^c, \bs{\theta}_{m^c}^{c})$. The algorithm proposes a move to $(m^p, \bs{\theta}_{m^p}^p)$, by drawing $(m^p, \bs{\theta}_{m^p}^p)$ from a proposal distribution  $q(m^p, \bs{\theta}_{m^p}^p|m^c, \bs{\theta}_{m^c}^c)$. This draw is accepted with probability
\begin{equation*}
    \alpha = \text{ min } \Bigg\{1, \frac{p(m^{p}, \bs{\theta}_{m^p}^{p}|\bs{x}) q(m^c, \bs{\theta}_{m^c}^{c}|m^p, \bs{\theta}_{m^p}^{p})}
    {p(m^{c}, \bs{\theta}_{m^c}^{c}|\bs{x}) q(m^p, \bs{\theta}_{m^p}^{p}|m^c, \bs{\theta}_{m^c}^{c})} \Bigg\},   
\end{equation*}
where $p(\cdot)$ is a target distribution, namely the product of the likelihood and the prior. The target and proposal distributions vary based on the type of move taken in the sampling scheme. The distribution $q(m^p, \bs{\theta}_{m^{p}}^{p}| m^c, \bs{\theta}_{m^{c}}^{c})$ is described as follows:
\begin{align*}
    q(m^p, \bs{\theta}_{m^p}^{p}|m^c, \bs{\theta}_{m^{c}}^{c}) = q(m^p|m^c)  q(\bs{\theta}_{m^p}^{p}| m^p, m^c, \bs{\theta}_{m^c}^{c}) 
    = q(m^p|m^c)  q(\bs{\xi^p_{m^p}}, \bs{\tau}_{m^p}^{2p}, \bs{\beta}_{m^p}^{p} | m^p, m^c, \bs{\theta}_{m^c}^c) \\
    = q(m^p|m^c)  q(\bs{\xi}_{m^p}^p|m^p, m^c, \bs{\theta}_{m^c}^c)  q(\bs{\tau}_{m^p}^{2p}|\bs{\xi}_{m^p}^p, m^p, m^c, \bs{\theta}_{m^c}^c)  q(\bs{\beta}_{m^p}^{p}|\bs{\tau}_{m^p}^{2p}, \bs{\xi}_{m^p}^p, m^p, m^c, \bs{\theta}_{m^c}^c).
\end{align*}
To draw $(m^p, \bs{\theta}_{m^p}^p)$, one must first draw $m^p$, and then $\bs{\xi}_{m^p}^p$, $\tau_{m^p}^{2p}, \text{ and } \bs{\beta}_{m^p}^{p}$. The number of segments $m^p$ is drawn from the proposal distribution $q(m^p|m^c)$. Let $M$ be the maximum number of segments and $m^{c}_{2,\text{min}}$ be the number of current segments containing at least $2  t_{min}$ data points. The proposal is as follows:
\begin{align*}
    q(m^p = k | m^c)= \left\{
                \begin{array}{ll}
                  1/2 \text{ if } k = m^c - 1, m^c + 1 \text{ and } m^{c} \neq 1, M, m_{2,\text{min}}^c \neq 0\\
                  1 \text{ if } k = m^{c}-1 \text{ and } m^{c} = M \text{ or } m_{2,\text{min}}^{c} = 0 \\
                  1 \text{ if } k = m^{c} + 1 \text{ and } m^{c} = 1
                \end{array}
              \right.
\end{align*}
Conditional on the proposed number of change points $m^p$, a new partition $\bs{\xi}_{m^p}^{p}$, a new vector of covariance amplitude parameters $\bs{\tau}_{m^p}^{2p}$, and a new vector of regression coefficients, $\bs{\beta}_{m^p}^{p}$ are proposed. $\tau^{2}$ is described as a smoothing parameter \cite{Rosen2012} or amplitude parameter.

Now, we describe the process of the \textbf{birth} of new segments, where the number of proposed change points increases. Suppose that $m^p = m^c + 1$. As defined before, a time series partition,
    \begin{align*}
    \bs{\xi}^{p}_{m^p} = (\xi^c_{0,m^c},...,\xi^c_{k^{*}-1,m^c},\xi_{k^{*},m^{p}}^{p}, \xi_{k^{*},m^{c}}^{c},...,\xi_{m^c,m^c}^{c})    
    \end{align*}
    is drawn from the proposal distribution $q(\bs{\xi}_{m^p}^{p}|m^p, m^c, \bs{\theta}_{m^c}^c)$. The algorithm first proposes a partition by selecting a random segment $j = k^{*}$ to split. Then, a random point $t^{*}$ within the segment $j=k^{*}$ is selected to be the proposed partition point. This is subject to a constraint, $\xi_{k^{*}-1, m^c}^{c} + t_{\text{min}} \leq t* \leq \xi_{k^{*},m^c}^c - t_{\text{min}}$. The proposal distribution is then computed as follows:
    \begin{align*}
        q(\xi_{j,m^p}^{p} = t^{*} | m^p, m^c, \bs{\xi}_{m^c}^{c}) =  p(j=k^{*} | m^p, m^c, \bs{\xi}_{m^c}^{c}); \\
         p(\xi_{k^{*}, m^p}^{p} = t^{*} | j=k^{*}, m^p, m^c, \bs{\xi}_{m^c}^c)
        =  \frac{1}{m_{2 \text{min}}^{c}(n_{k^{*}, m^c} - 2t_{\text{min}}+1)}.
    \end{align*}
Then, the vector of amplitude parameters
    \begin{align*}
        \tau_{m^p}^{2p} = (\tau_{1,m^c}^{2c},...,\tau_{k^{*}-1,m^c}^{2c},  \tau_{k^{*},m^p}^{2p}, \tau_{k^{*}+1,m^p}^{2p}, \tau_{k^{*}+1,m^c}^{2c},...,\tau_{m^c, m^c}^{2c})
    \end{align*}
    is drawn from the proposal distribution $q(\bs{\tau}_{m^p}^{2p}|m^p, \bs{\xi}_{m^p}^p, m^c, \bs{\theta}_{m^c}^c) = q(\bs{\tau}_{m^p}^{2p}|m^p, \bs{\tau}_{m^c}^{2c}).$ The algorithm is based on the RJMCMC algorithm of \cite{GREEN1995}. This draws from a uniform distribution $u \sim U[0,1]$ and defines $\tau_{k^{*}, m^p}^{2p}$ and $\tau_{k^{*}+1, m^p}^{2p}$ in terms of $u$ and $\tau_{k^{*}, m^c}^{2c}$ as follows:
    \begin{align}
    \label{eq:birth1}
        \tau_{k^{*, m^p}}^{2p} =   \frac{u}{1-u}\tau_{k^{*}, m^c}^{2c}; \\
        \tau_{k^{*}+1, m^p}^{2p} =   \frac{1-u}{u}\tau_{k^{*}, m^c}^{2c}.
        \label{eq:birth2}
    \end{align}
Then, the vector of coefficients
    \begin{align*}
        \bs{\beta}_{m^p}^p = (\bs{\beta}_{1,m^c}^{c},...,\bs{\beta}_{k^{*}-1,m^c}^{c},  \bs{\beta}_{k^{*}, m^p}^p, \bs{\beta}_{k^{*}+1,m^p}^{p}, \bs{\beta}_{k^{*}+1,m^c}^{c},...,\bs{\beta}_{m^c,m^c}^{c})
    \end{align*}
    is drawn from the proposal distribution  $q(\bs{\beta}_{m^p}^p|\bs{\tau}_{m^p}^{2p},\bs{\xi}_{m^p}^{2p},m^p, m^c, \bs{\theta}_{m^c}^c) = q(\bs{\beta}_{m^p}^{p}|\bs{\tau}_{m^p}^{2p}, \bs{\xi}_{m^p}^p, m^p)$. The vectors $\bs{\beta}_{k^{*}, m^p}^p$ and $\bs{\beta}_{k^{*}+1, m^p}^p$ are drawn from Gaussian approximations to the respective posterior conditional distributions $p(\bs{\beta}_{k^{*}, m^p}^p|\bs{x}_{k^{*}}^p, \tau_{k^{*}, m^p}^{2p}, m^p)$ and  $p(\bs{\beta}_{k^{*}+1, m^p}^{p}|\bs{x}_{k^{*}+1}^p, \tau_{k^{*}+1, m^p}^{2p}, m^p)$, respectively. Here, $\bs{x}_{k^{*}}^p$ and $\bs{x}_{k^{*}+1}^p$ refer to the subsets of the time series across segments $k^{*}$ and $k^{*}+1$, respectively. Then, $\bs{\xi}_{m^p}^p$  determines $\bs{x_{*}}^p = (\bs{x}_{k^{*}}^{p'}, \bs{x}_{k^{*}+1}^{p'})'$. We provide an example for illustration: the coefficient $\bs{\beta}_{k^{*}, m^p}^{p}$ is drawn from the Gaussian distribution $N(\bs{\beta}_{k^{*}}^{\text{max}}, \Sigma_{k^{*}}^{\text{max}})$, where  
    \begin{align*}
     \bs{\beta}_{k^{*}}^{\text{max}}=    \argmax_{\bs{\beta}_{k^{*}, m^p}^{p}} p(\bs{\beta}^p_{k^{*}, m^p}|\bs{x}_{k^{*}}^p, \tau_{k^{*}, m^p}^{2p}, m^p)
    \end{align*} and 
    \begin{align*}
      \Sigma_{k^{*}}^{\text{max}} =-\Bigg \{ \pdv{\log p(\bs{\beta}_{k^{*}, m^p}^p | \bs{x}_{k^{*}}^p, \tau_{k^{*}, m^p}^{2p}, m^p)}{\bs{\beta}_{k^{*}, m^p}^{p}}{\bs{\beta}_{k^{*}, m^p}^{p'}} \Bigg|_{\bs{\beta}_{k^{*}, m^p}^{p} = \bs{\beta}_{k^{*}}^{\text{max}}}  \Bigg \}_.^{-1}
    \end{align*}
    For the birth move (that is the increase in $m$), the probability of acceptance is $\alpha = \min\{1,A\}$, where $A$ is equal to
    \begin{align*}
        \Bigg| \pdv{(\tau_{k^{*}, m^p}^{2p}, \tau_{k^{*}+1, m^p}^{2p})}{(\tau_{k^{*}, m^c, u}^{2c})}  \Bigg|\frac{p(\bs{\theta}_{m^p}^p|\bs{x}, m^p) p(\bs{\theta}_{m^p}^p|m^p)p(m^p)}{p(\bs{\theta}_{m^p}^p|\bs{x}, m^p) p(\bs{\theta}_{m^c}^c|m^c)p(m^c)}  \frac{p(m^{c}|m^p)p(\bs{\beta}_{k^{*}, m^c}^{c})}{p(m^p|m^c)p(\xi_{k^{*}, m^p}^{m^p}|m^p, m^c) p(u) p(\bs{\beta}^p_{k^{*}, m^p})p(\bs{\beta}_{k^{*}+1,m^{p}}^{p})}.  
    \end{align*}
    Above, $p(u) = 1, 0 \leq u \leq 1,$ while $p(\bs{\beta}_{k^{*}, m^p}^{p})$ and $p(\bs{\beta}_{k^{*}+1, m^p}^{p})$ are the density functions of Gaussian proposal distributions $N(\bs{\beta}_{k^{*}}^{\text{max}}, \Sigma_{k^{*}}^{\text{max}})$ and $N(\bs{\beta}_{k^{*}+1}^{\text{max}}, \Sigma_{k^{*}+1}^{\text{max}})$, respectively. The above Jacobian is computed as
    \begin{align*}
        \bigg| \pdv{(\tau_{k^{*}, m^p}^{2p}, \tau_{k^{*}+1, m^p}^{2p})}{(\tau^{2c}_{k^{*}, m^c}, u)} \bigg| = \frac{2 \tau_{k^{*}m^c}^{2c}}{u(1-u)} = 2(\tau_{k^{*}, m^p}^{p} + \tau_{k^{*}+1, m^p}^{p})^{2}.
    \end{align*}

Now, we move to describe the process of the \textbf{death} of new segments. This is where the number of proposed change points will decrease, or $m^p = m^c - 1$. A time series partition is denoted
\begin{align*}
\bs{\xi}_{m^p}^{p} = (\xi_{0,m^c}^{c},...,\xi_{k^{*}-1,m^c}^{c}, \xi_{k^{*}+1,m^c}^{c},...,\xi_{m^c,m^c}^{c}),
\end{align*}
is proposed by randomly selecting a single change point from $m^c - 1$ candidates, and then removing it. The change point selected for the removal is denoted $j=k^{*}$. There are $m^c -1$ possible change points available for removal among the $m^c$ segments currently in existence in the model. The proposal can choose each change point with equal probability, that is,
\begin{align*}
    q(\xi_{j, m^p}^p|m^p, m^c, \bs{\xi}_{m^c}^c) = \frac{1}{m^c - 1}.
\end{align*}
The updated vector of amplitude parameters 
\begin{align*}
\bs{\tau}_{m^p}^{2p} = (\tau_{1, m^c}^{2c},...,\tau_{k^{*}-1,m^c}^{2c},\tau_{k^{*},m^p}^{2c}, \tau_{k^{*}+2,m^c}^{2c},...,\tau_{m^c,m^c}^{2c})    
\end{align*}
 is then drawn from the proposal distribution  $q(\bs{\tau}_{m^p}^{2p}|m^p, \bs{\xi}_{m^p}^p, m^c, \bs{\theta}_{m^c}^{c}) = q(\bs{\tau}_{m^p}^{2p}| m^p, \bs{\tau}_{m^c}^{2c})$. Then, one amplitude parameter $\tau_{k^{*}, m^p}^{2p}$ is formed from two candidate amplitude parameters, $\tau_{k^{*},m^c}^{2c}$ and $\tau_{k^{*}+1,m^c}^{2c}$,. by reversing the equations (\ref{eq:birth1}) and (\ref{eq:birth2}). Specifically,
\begin{align*}
    \tau_{k^{*}, m^p}^{2p} = \sqrt{\tau_{k^{*}, m^{c}}^{2c} \tau_{k^{*}+1, m^c}^{2c}}.
\end{align*}
Finally, the updated vector of regression coefficients,
\begin{align*}
    \bs{\beta}_{m^p}^{p} = (\beta_{1,m^c}^{c},...,\beta_{k^{*}-1,m^c}^{c}, \beta_{k^{*}, m^p}^{p}, \beta_{k^{*}+2,m^c}^{c},...,\beta_{m^c,m^c}^{c})
\end{align*}
is drawn from the proposal distribution  $q(\bs{\beta}_{m^p}^p|\bs{\tau}_{m^p}^{2p}, \bs{\xi}_{m^p}^p, m^p, m^c, \theta_{m^c}^c) = q(\bs{\beta}_{m^p}^{p}|\bs{\tau}_{m^p}^{2p}, \bs{\xi}_{m^p}^p, m^p)$. The vector of regression coefficients is drawn from a Gaussian approximation to the posterior distribution $p(\beta_{k^{*},m^p}|\bs{x}, \tau_{k^{*}, m^p}^{2p}, \bs{\xi}^p_{m^p}, m^p)$ with the same underlying procedure for the vector of coefficients within the birth step. The probability of an acceptance is the inverse of the corresponding birth step. If the move is accepted then the following updates of the current values occur: $m^c=m^p$ and $\bs{\theta}_{m^c}^c = \bs{\theta}_{m^p}^{p}$.

Finally, we describe the \textbf{within-model moves:}
from this point on, the number of change points $m$ will be fixed and the notation describing the dependence on the number of change points (explicitly) will be removed. There are two sub-components of the within-model move. In the first piece, the change points could be relocated, and conditional on this relocation, the coefficients of the candidate basis functions will be updated. These steps are jointly accepted or rejected within a Metropolis-Hastings step, and the amplitude parameters are also updated in a separate Gibbs sampling step. 

We assume the chain is located at $\bs{\theta}^{c} = (\bs{\xi}^{c}, \bs{\beta}^{c})$. The proposal $\bs{\theta}^p = (\bs{\xi}^p, \bs{\beta}^p)$ is as follows: first, a change point $\xi_{k^{*}}$ is selected for potential relocation from $m-1$ candidate change points. Next, a position within the interval $[\xi_{k^{*}-1}, \xi_{k^{*}+1}]$ is chosen, subject to the constraint that the new location will be at least $t_{\text{min}}$ data points away from $\xi_{k^{*}-1}$ and $\xi_{k^{*}+1}$, so that
\begin{align*}
    \Pr(\xi^p_{k^{*}}=t) = \Pr (j=k^{*})  \Pr (\xi_{k^{*}}^{p}=t|j=k^{*}), 
\end{align*}
where $\Pr(j=k^{*}) = (m-1)^{-1}$. A mixture distribution for $\Pr(\xi_{k^{*}}^p=t|j=k^{*})$ is constructed to explore the space efficiently, so
\begin{align*}
    \Pr(\xi_{k^{*}}^{p}=t|j=k^{*}) =  \pi q_1 (\xi_{k^{*}}^p = t| \xi_{k^{*}}^{c}) + (1-\pi) q_2 (\xi_{k^{*}}^p=t|\xi_{k^{*}}^c),
\end{align*}
where $q_1(\xi_{k^{*}}^p = t| \xi_{k^{*}}^c) = (n_{k^{*}} + n_{k^{*}+1}-2t_{\text{min}} + 1)^{-1}, \xi_{k^{*}-1} + t_{\text{min}} \leq t \leq \xi_{k^{*}+1} - t_{\text{min}}$ and 

\begin{align*}
    q_2(\xi_{k^{*}}^p = t|\xi_{k^{*}}^{c})= 
\left\{
            \begin{array}{ll}
              0 \text{ if } |t-\xi^c_{k^{*}}| > 1;  \\
            1/3 \text{ if } |t-\xi^c_{k^{*}}| \leq 1, n_{k^{*}} \neq t_{\text{min}} \text{ and } n_{k^{*}+1} \neq t_{\text{min}};  \\ 
            1/2 \text{ if } t-\xi_{k^{*}}^{c} \leq 1, n_{k^{*}} = t_{\text{min}} \text{ and } n_{k^{*}+1} \neq t_{\text{min}}; \\
            1/2 \text{ if } \xi_{k^{*}}^{c} - t \leq 1, n_{k^{*}} \neq t_{\text{min}} \text{ and } n_{k^{*}+1} = t_{\text{min}}; \\
            1 \text{ if } t = \xi_{k^{*}}^{c}, n_{k^{*}} = t_{\text{min}} \text{ and } n_{k^{*}+1} = t_{\text{min}}.
            \end{array}
          \right.
\end{align*}
The support of $q_1$ has $n_{k^{*}} + n_{k^{*}+1} - 2t_{\text{min}} + 1$ points while $q_2$ has at most three. The term $q_2$ alone may result in a meaningfully higher acceptance rate for the M-H step, but it will slowly explore the parameter space. This issue is exacerbated in higher dimensional settings. The $q_1$ component will allow for larger steps, and can produce a compromise between higher acceptance rates and a deep exploration of parameter space.

 Then, $\bs{\beta^{p}_{j}}, j=k^{*}, k^{*}+1$ is drawn from an approximation to $\prod^{k^{*}+1}_{j=k^{*}} p(\bs{\beta}_j|\bs{x}_j^p, \tau_j^{2})$, following the corresponding step in the between-model move. The proposal distribution, evaluated at $\bs{\beta}^{p}_j, j=k^{*}, k^{*}=1$, is
\begin{align*}
    q(\bs{\beta}_{*}^{p}|\bs{x}_{*}^p, \bs{\tau}_{*}^{2}) = \prod^{k^{*}+1}_{j=k^{*}} q(\bs{\beta}_{j}^p|\bs{x}_j^p, \tau_j^{2}),
\end{align*}
where $\bs{\beta}_{*}^p = (\bs{\beta}^{p'}_{k^{*}}, \bs{\beta}^{p'}_{k^{*}+1})'$ and $\bs{\tau}_{*}^{2} = (\tau^{2}_{k^{*}}, \tau^{2}_{k^{*}+1})'$. This proposal distribution is also evaluated at current values of $\bs{\beta}_{*}^{c} = (\beta^{c'}_{k^{*}}, \beta^{c'}_{k^{*}+1})'$. $\beta_{*}^p$ is then accepted with probability
\begin{equation*}
    \alpha = \min \Bigg\{ 1, \frac{p(\bs{x}_{*}^p|\bs{\beta}_{*}^{p}) p(\bs{\beta}_{*}^{p}|\bs{\tau}_{*}^{2}) q(\bs{\beta}_{*}^{c}|\bs{x}^{c}_{*}, \bs{\tau}_{*}^{2})} {p(\bs{x}_{*}^c|\bs{\beta}_{*}^{c}) p(\bs{\beta}_{*}^{c}|\bs{\tau}_{*}^{2}) q(\bs{\beta}_{*}^{p}|\bs{x}^{p}_{*}, \bs{\tau}_{*}^{2})} \Bigg\},
\end{equation*}
where $\bs{x}_{*}^{c} = (\bs{x}^{c'}_{k^{*}},\bs{x}^{c'}_{k^{*}+1})$. When the draw is accepted, we update the partition and regression coefficients $(\xi^{c}_{k^{*}}, \beta_{*}^{c}) = (\xi^{p}_{k^{*}}, \beta_{*}^{p})$.
Finally, we draw $\tau^{2p}$ from 
\begin{align*}
    p(\tau_{*}^{2}|\bs{\beta}_{*}) = \prod^{k^{*}+1}_{j=k^{*}} p(\tau_j^{2}|\beta_j).
\end{align*}
This is a Gibbs sampling step, and as such the draw is accepted with probability 1.



\bibliographystyle{_elsarticle-num-names}
\bibliography{__References.bib}
\end{document}